\documentclass[onecolumn,12pt,draftclsnofoot]{IEEEtran}

\usepackage{mathrsfs}
\usepackage{mathptmx}
\usepackage{amsmath}
\usepackage{amsfonts}
\usepackage{amssymb}
\usepackage{graphicx}
\usepackage{xcolor}
\usepackage{float}
\usepackage{subfigure}
\usepackage{psfrag}
\usepackage{bm}
\usepackage{cite}

\newtheorem{defi}{Definition}
\newtheorem{thm}{Theorem}

\newtheorem{cor}{Corollary}
\newtheorem{rem}{Remark}

\newtheorem{lem}{Lemma}
\def\tp{\mathrm{T}}

\def\Ds{\displaystyle}
\def\tp{\mathrm{T}}
\def\Ds{\displaystyle}
\def\Z{\mathbb{Z}}
\def\R{\mathbb{R}}
\def\P{\mathbb{P}}
\def\*{\star}

\def\L{\mathcal{L}}
\def\F{\mathbb{F}}

\renewcommand\baselinestretch{1.25}

\begin{document}

\title{\bf\huge Fundamental Trackability Problems for Iterative Learning Control}

\author{Deyuan Meng, {\it Senior Member}, {\it IEEE} and Jingyao Zhang
\thanks{The authors are with the Seventh Research Division, Beihang University (BUAA), Beijing 100191, P. R. China, and also with School of Automation Science and Electrical Engineering, Beihang University (BUAA), Beijing 100191, P. R. China (e-mail: dymeng@buaa.edu.cn).}
}

\date{}
\maketitle

\begin{abstract}
Generally, the classic iterative learning control (ILC) methods focus on finding design conditions for repetitive systems to achieve the perfect tracking of any specified trajectory, whereas they ignore a fundamental problem of ILC: whether the specified trajectory is trackable, or equivalently, whether there exist some inputs for the repetitive systems under consideration to generate the specified trajectory? The current paper contributes to dealing with this problem. Not only is a concept of trackability introduced formally for any specified trajectory in ILC, but also some related trackability criteria are established. Further, the relation between the trackability and the perfect tracking tasks for ILC is bridged, based on which a new convergence analysis approach is developed for ILC by leveraging properties of a functional Cauchy sequence (FCS). Simulation examples are given to verify the effectiveness of the presented trackability criteria and FCS-induced convergence analysis method for ILC.
\end{abstract}


\section{Introduction}\label{sec1}

\IEEEPARstart{I}{terative} learning control (ILC) is proposed for the class of robots executing repetitive tasks, which aims at bettering the execution performances of robots for the current operation (trial or iteration) by taking advantage of the saved information from the past operations \cite{akm:84}. Because of its ability of achieving high-precision tracking tasks, ILC has been well developed for the past three decades and successfully applied in many fields, such as flexible structures \cite{hmhg:18}, railway traffic systems \cite{yh:20}, batch processes \cite{ll:07}, and network systems \cite{mz:21}. For more explanations of ILC, the readers are referred to the surveys of, e.g., \cite{bta:06,acm:07,x:11}. It is worth highlighting that due to the salient two-dimensional (2-D) operating rules with respect to the independent time and iteration axes, classic ILC methods require limited information of the controlled system, employ simple (mostly, the PID-type) controller structures, and are easy-to-implement by resorting to distinct convergence analysis strategies from typical feedback-based control methods (generally, via the contraction mapping and fixed point theorems instead of the Lyapunov theories).

In the classic ILC framework, ILC is generally implemented for the controlled system to track any specified (or desired) trajectory perfectly on a finite time interval through the following typical steps (see also \cite{bta:06,acm:07,x:11}):
\begin{enumerate}
\item[S1)]
making necessary assumptions on the controlled system, especially those on its dynamics repetitiveness, identical initial alignment condition, and system relative degree; 

\item[S2)]
choosing which type of ILC updating laws is employed, or mostly, deciding what kind of PID-type ILC updating laws needs to be adopted;

\item[S3)]
finding ILC design conditions to ensure the convergence of the resulting iterative process, especially by constructing contraction mapping conditions for the ILC process.
%
\end{enumerate}

\noindent Although many remarkable results have been established in the ILC framework developed by the steps S1)--S3), a fundamental {\it trackability} problem of ILC remains beyond this framework:
\begin{enumerate}
\item[P1)]
whether the specified trajectory is {\it trackable} in ILC, or in other words, are there any ILC updating laws driving the controlled system to generate the specified trajectory?
\end{enumerate}

Regarding the fundamental problem P1), seldom ILC results have been reported to answer it. Instead it is common in classic ILC to directly assume that the specified trajectory is {\it realizable} (namely, there exists a {\it unique} input for the controlled system to generate the specified trajectory) \cite{akm:84,s:95,ack:93}. By contrast, trackability is obviously a more available property for ILC than realizability by avoiding imposing the uniqueness requirement. Of particular note is that not only can the realizability-induced ILC results be not utilized to address the fundamental problem P1), but also the need of the realizability assumption may limit the application range of ILC significantly. It has been disclosed recently in \cite{mw:21} that there exists a large class of ILC problems, where the specified trajectories are trackable but not realizable. In such ILC problems, those ILC analysis methods and results established with the realizability assumption naturally does not work any longer. Conversely, the realizability can be addressed as a special case of the trackability for ILC \cite{mw:21}. Consequently, the solving of the fundamental problem P1) not only is crucial for ILC but also may bring novel insights into its development. This observation has been verified for the case of discrete-time ILC in \cite{mw:21}, where trackability provides exactly the necessary and sufficient guarantee for the ILC perfect tracking tasks and, moreover, makes it possible to connect ILC with controllability of discrete systems. Unfortunately, the results presented in \cite{mw:21} are thanks to the lifting technique induced by the discrete-time characteristic of the ILC system, which thus can not be applied to continuous-time ILC. In fact, the fundamental problem P1) has not been considered for ILC in the presence of continuous-time systems to the best of our knowledge.

Another benefit of the realizability assumption is to induce a class of convergence analysis methods for ILC, which is called the {\it indirect method}, since it does not directly contribute to the accomplishment of the tracking task. Thanks to the assumption on the existence and uniqueness of the input for the controlled system to output the specified trajectory, the indirect method is possible and convenient to first obtain the convergence analysis of the resulting input sequence along the iteration axis, through which the primary tracking task can then be indirectly achieved for ILC by taking full advantage of the time-domain dynamics of the controlled system (see also \cite{akm:84,s:95,ack:93}). In comparison with the indirect method, there exists a class of {\it direct methods} for the convergence analysis of ILC, which is directly devoted to realizing the tracking task without imposing the realizability assumption (see, e.g., \cite{mm:17a,zm:21tcyb,mz:20tnnls}). However, the direct method of ILC generally does not devote much attention to the iterative evolution process of the input, where it is unclear how the ILC updating law of input works in ensuring the controlled system to accomplish the tracking task. Moreover, even contradictory assumptions upon the controlled system and conditions for the ILC design are produced when applying the direct and indirect methods to the same ILC problem for multi-input multi-output (MIMO) systems. It is actually because of the limitation of the existing analysis methods in ILC, where the direct and indirect methods are implemented from the dual perspectives of output and input of the controlled systems, respectively. To the best of our knowledge, the following fundamental problem concerned with the convergence analysis methods of ILC is unaddressed:
\begin{enumerate}
\item[P2)]
whether and how can new convergence analysis methods for continuous-time ILC be established to avoid resulting in contradictory assumptions or conditions when dealing with  ILC convergence even from the perspectives of both output and input of the MIMO controlled systems?
\end{enumerate}

In this paper, we are devoted to coping with the fundamental ILC problems P1) and P2) with a focus on MIMO, continuous-time linear systems. Because they are distinct problems of ILC owing to the 2-D dynamics process of ILC, the well-developed analysis and design results for feedback-based control methods do not work any more, which is particularly true when it comes to ILC for continuous-time systems due to the resulting hybrid 2-D discrete and continuous dynamics \cite{pbt:19}. Despite this issue, we leverage the properties of polynomial matrix and functional Cauchy sequence (FCS) to establish a new framework for ILC, in which we can successfully hanlde the fundamental problems P1) and P2). In comparison with the existing ILC literature, the main contributions for our paper are summarized as follows.
\begin{enumerate}
\item[1)]
We formally introduce a definition of trackability for any specified trajectory in ILC by resorting to the frequency-domain algebraic equations. Furthermore, we explore the trackability criteria for ILC systems by taking advantage of the polynomial matrix properties. It is also shown that both the system relative degree and the initial alignment condition have great influences on whether the specified trajectory is trackable in ILC. In addition, this provides a strong explanation about why the system relative degree and the initial alignment condition are the fundamentally required assumptions in classic ILC from the trackability viewpoint of the specified trajectory.

\item[2)]
We propose a general feedback-based design method for ILC updating laws in the presence of any tracking tasks. Under the trackability premise of the specified trajectory, our proposed method closely connects the design of ILC updating laws with a class of state feedbacks constructed in the iteration domain. This is thanks to generalizing the design idea of \cite{mw:21} with the frequency-domain methods, which also narrows the gap between the design of classic continuous-time ILC and that of feedback-based control methods. In particular, our design method collapses into providing PID-type ILC updating laws with appropriate selections of the gain function matrix.

\item[3)]
We develop an FCS-induced method for the convergence analysis of ILC, through which we can leverage a unified design condition to achieve the ILC convergence for the MIMO controlled systems from the perspectives of both output and input. Particularly, it is shown that the steady-state input obtained after an ILC process depends heavily on the initial input. Moreover, we bridge the relationship between the trackability of a trajectory specified for ILC and the accomplishment of the resulting perfect tracking objective. More specifically, we reveal that regardless of whether the MIMO controlled systems are over-actuated or under-actuated, the perfect tracking objectives for ILC can be accomplished under certain ILC updating laws if and only if the specified trajectories are trackable in ILC.
\end{enumerate}

In addition, our developed results can contribute to bettering the typical steps S1)--S3) of classic ILC. We verify the validity of them through two simulation examples considered for over-actuated and under-actuated systems, respectively.

The rest of this paper is organized as follows. In Section \ref{sec2}, the trackability problem for continuous-time ILC is introduced. The trackability criteria and the FCS-induced tracking analysis of ILC are established in Sections \ref{sec6} and \ref{sec3}, respectively. Two simulation examples are provided in Section \ref{sec4}, and finally, the conclusions are made in Section \ref{sec5}.

{\it Notations:} For any $T>0$, let $C_{n}[0,T]$ (respectively, $C^{1}_{n}[0,T]$) be the space of $n$-dimensional real-valued vector functions that are continuous (respectively, continuously differentiable) on an interval $[0,T]\triangleq\{t\in\mathbb{R}|0\leq t\leq T\}$. Given $f(t)\in\mathbb{R}^{n}$, $\forall t\in[0,T]$, let $\|f(t)\|$ be any vector norm of it (see, e.g.,\cite[p. 265]{hj:85} for $l_{\infty}$ norm and $l_{p}$, $p\geq1$ norm), based on which its $\lambda$-norm ($\lambda>0$) is defined as $\|f(t)\|_{\lambda}=\sup_{t\in[0,T]}\left(\|f(t)\|e^{-\lambda t}\right)$. The Laplace transform of $f(t)$ is denoted as $F(s)=\mathcal{L}\left[f(t)\right]$, for which the inverse Laplace transformation writes as $f(t)=\mathcal{L}^{-1}\left[F(s)\right]$. In addition, let $\R\P^{q\times p}(s)$ be the set of $q\times p$ polynomial matrices (i.e., those $q\times p$ matrices whose elements are polynomials in $s$ with the real coefficients) \cite{am:06}, and particularly for $q=1$ and $p=1$, let it be denoted as $\R\P(s)$. With this fact, let $\R\F^{q\times p}(s)$ be the set of $q\times p$ rational fraction matrices, i.e.,
\[\aligned
\R\F^{q\times p}(s)\triangleq\Bigg\{G(s)=\left[g_{ij}(s)\right]\big|g_{ij}(s)=\frac{\Ds n_{ij}(s)}{\Ds d_{ij}(s)},&\\
\hbox{with}~n_{ij}(s),d_{ij}(s)\in\R\P(s),d_{ij}(s)\not\equiv0,&\\
\forall i=1,2,\cdots,q,j=1,2,\cdots,p&\Bigg\}.
\endaligned\]

\noindent For any $G(s)\in\R\F^{q\times p}(s)$ satisfying $\lim_{s\to\infty}G(s)=D$ for some constant matrix $D\in\R^{q\times p}$, if $D\neq0$ (respectively, $D=0$), then it is said to be proper (respectively, strictly proper) \cite{am:06}, where it is alternatively called a proper (respectively, strictly proper) transfer function matrix for convenience. Let $\delta(t)$ be the unit impulse function \cite{am:06,lmt:07}, namely, we have $\mathcal{L}\left[\delta(t)\right]=1$.

Denote $\mathbb{Z}\triangleq\{1,2,3,\cdots\}$ and $\mathbb{Z}_{+}\triangleq\{0,1,2,\cdots\}$ as the sets of the positive and nonnegative integers, respectively. For $j\in\mathbb{Z}_{+}$, let $f^{(j)}(t)$ be the $j$th-order derivative of any function $f(t)$, i.e., $f^{(j)}(t)\triangleq d^{j}f(t)/dt^{j}$. If $f^{(j)}(t)$ exists for any $j\in\mathbb{Z}_{+}$, then $f(t)$ is called a smooth function. For any sequence $\left\{\xi_{j}:j\in\mathbb{Z}_{+}\right\}$, we denote $\sum_{j=0}^{-1}\xi_{j}=0$ and define a forward difference operator such that $\Delta:\xi_{j}\to\Delta\xi_{j}=\xi_{j+1}-\xi_{j}$, $\forall j\in\mathbb{Z}_{+}$.

\section{Problem Statement}\label{sec2}

Consider a continuous-time MIMO ILC system running on a finite-time interval, denoted by $t\in[0,T]$, and along an iteration axis, denoted by $k\in\mathbb{Z}_{+}$. If the output and control input for this system are, respectively, denoted by $y_{k}(t)\in\mathbb{R}^{q}$ and $u_{k}(t)\in\mathbb{R}^{p}$, then the objective of ILC is generally realized in the sense that the input $u_{k}(t)$ with some updating laws can be designed along the iteration axis $k\in\mathbb{Z}_{+}$ to make the output $y_{k}(t)$ able to arrive at the perfect tracking of a desired output trajectory $y_{d}(t)\in\mathbb{R}^{q}$ specified over $[0,T]$ from the beginning to the end, namely,
\begin{equation}\label{b1}
\lim_{k\to\infty}y_{k}(t)=y_{d}(t),\quad\forall t\in[0,T].
\end{equation}

\noindent For this perfect tracking task of ILC, the fundamental problem P1) naturally arises: whether $y_{d}(t)$ is trackable, or equivalently, whether there exists some desired input, denoted as $u_{d}(t)\in\mathbb{R}^{p}$, to correspondingly generate $y_{d}(t)$? However, it is subject to the lack of consideration for the fundamental trackability problem P1) in ILC \cite{mw:21}, where there even do not exist any trackability-related concepts, properties, methods, or results that have been introduced formally and clearly in the presence of continuous-time ILC systems to our knowledge. 

To clearly explore the fundamental trackability problem P1), let $Y_{k}(s)=\mathcal{L}\left[y_{k}(t)\right]$ and $U_{k}(s)=\mathcal{L}\left[u_{k}(t)\right]$, and then we focus specifically on the linear system given in the frequency-domain form of
\begin{equation}\label{b2}
\aligned
Y_{k}(s)&=G_{1}(s)U_{k}(s)+G_{2}(s)D(s)\\
\endaligned
\end{equation}

\noindent where $D(s)\triangleq\mathcal{L}\left[d(t)\right]$ with $d(t)\in\mathbb{R}^{m}$ to represent the possible additional inputs, such as the disturbance (noise) and the initial (output or state) condition, and $G_{1}(s)\in\R\F^{q\times p}(s)$ and $G_{2}(s)\in\R\F^{q\times m}(s)$ are two transfer function matrices. For any specified trajectory $y_{d}(t)$ of the system (\ref{b2}), we consider the general case that its Laplace transform exists, and then let $Y_{d}(s)=\mathcal{L}\left[y_{d}(t)\right]$.

We now present a formal concept of trackability in ILC from the perspective of solving algebraic equations by incorporating the advantage of the Laplace transformation.

\begin{defi}\label{defi2}
For the system (\ref{b2}), a specified output trajectory $y_{d}(t)\in\R^q$ is called {\it trackable} in ILC if there exists some input $u_{d}(t)\in\R^p$ such that $U_{d}(s)=\L\left[u_d(t)\right]$ satisfies the frequency-domain algebraic equation:
\begin{equation}\label{b3}
G_{1}(s)U_{d}(s)=Y_{d}(s)-G_{2}(s)D(s).
\end{equation}

\noindent Particularly, if the algebraic equation (\ref{b3}) has a {\it unique} solution, then $y_{d}(t)$ is called {\it realizable} in ILC.
\end{defi}

For Definition \ref{defi2}, the solving of the algebraic equation (\ref{b3}) is crucial, where the theory of polynomial matrices \cite{am:06} is useful.
In classic ILC, the realizability is a usually adopted assumption for the accomplishment of the tracking tasks (see, e.g., \cite{akm:84,s:95,ack:93}). However, Definition \ref{defi2} suggests that the realizability may not be required by ILC tracking tasks, whereas the trackability is necessarily needed. This is owing to avoiding the uniqueness requirement in the trackability, for which the realizability can actually be included as a trivial case of the trackability for ILC. We thus aim at dealing with the more fundamental trackability-related ILC problems, as presented below.

{\it Problem statement.} For the system (\ref{b2}), the ILC problem that we address is to first determine whether the specified trajectory $y_{d}(t)$ is trackable and then design updating laws to accomplish the tracking task (\ref{b1}) in the presence of any trackable $y_{d}(t)$. To proceed, we further address how to get all inputs that generate the trackable $y_{d}(t)$ for the system (\ref{b2}). Of our additional interest is the robustness problem of our trackability-based ILC results with respect to iteration-varying uncertainties. 

We also introduce a new FCS-induced analysis approach to address the aforementioned trackability-related ILC problems. By directly focusing on the sequence of inputs generated from the proposed ILC updating law, we aim at exploring properties of the FCS to implement the ILC convergence analysis. Thanks to the implementation of FCS-based ILC analyses, we not only aim to avoid imposing some restrictive assumptions commonly needed in ILC, such as realizability and repetitiveness, but also arrive at unified design conditions to realize the convergence of ILC from the perspectives of both output and input, regardless of over-actuated or under-actuated MIMO controlled systems. This contributes to dealing with the fundamental problem P2) of the ILC convergence analysis.

Before proceeding further with exploring the given problem, we introduce a definition for an FCS, together with preliminary lemmas for (strictly) proper transfer function matrices.

\begin{defi}\label{defi1}
For any function $f_k(t)\in\mathbb{R}^{n}$, $\forall t\in[0,T]$, $\forall k\in \Z_+$, the resulting functional sequence $\{f_k(t):k\in \Z_+\}$ is called an FCS if, for any $\epsilon>0$, there exists some $N(\epsilon)\in\mathbb{Z}$ (i.e., $N(\epsilon)$ depends on $\epsilon$) such that $\left\|f_i(t)-f_j(t)\right\|_{\lambda}\leq\epsilon$, $\forall i,j\geq N(\epsilon)$.
\end{defi}

By Definition \ref{defi1}, an FCS refers to a functional sequence that satisfies the Cauchy criterion for the uniform convergence (see also \cite[Chapter 1, Theorem 5.3]{am:06}). In view of this observation, we propose a lemma to provide a guarantee for how to make a functional sequence generated by any proper transfer function matrix be an FCS.

\begin{lem}\label{lem08}
For any function $f_{k}(t)\in\mathbb{R}^{n}$, $\forall t\in[0,T]$, $\forall k\in\mathbb{Z}_{+}$, let $F_{k}(s)\triangleq\mathcal{L}\left[f_{k}(t)\right]$ be such that, for some $G_{F}(s)\in\R\F^{n\times n}(s)$,
\begin{equation*}\label{a13}
F_{k+2}(s)-F_{k+1}(s)=G_{F}(s)\left[F_{k+1}(s)-F_{k}(s)\right],\quad\forall k\in\mathbb{Z}_{+}.
\end{equation*}

\noindent Then the following three statements are equivalent:
\begin{enumerate}
\item
the sequence $\left\{f_{k}(t):k\in\mathbb{Z}_{+}\right\}$ is an FCS;

\item
the functional sequence $\left\{f_{k}(t):k\in\mathbb{Z}_{+}\right\}$ converges uniformly to some function $f_{\infty}(t)\in\mathbb{R}^{n}$ on $[0,T]$;

\item
$G_{F}(s)$ is proper such that
\begin{equation*}\label{a14}
\rho\left(\lim_{s\to\infty}G_{F}(s)\right)<1.
\end{equation*}
\end{enumerate}

\noindent Further, if $f_{k}(t)\in C_{n}[0,T]$, $\forall k\in\mathbb{Z}_{+}$, then $f_{\infty}(t)\in C_{n}[0,T]$.
\end{lem}

In addition to Lemma \ref{lem08}, the following one develops a closed property for the space of continuous functions under the action of proper transfer function matrices.

\begin{lem}\label{lem07}
If $\overline{G}(s)\in\R\F^{m\times n}(s)$ is proper, then for any $f(t)\in C_{n}[0,T]$, $\overline{f}(t)\in C_{m}[0,T]$ holds, where $\overline{f}(t)\triangleq\mathcal{L}^{-1}\left[\overline{G}(s)F(s)\right]$ with $F(s)\triangleq\mathcal{L}\left[f(t)\right]$. Moreover, if $\overline{G}(s)$ is strictly proper, then for any $v\in\mathbb{R}^{n}$, $\mathcal{L}^{-1}\left[\overline{G}(s)v\right]\in C_{m}[0,T]$ holds.
\end{lem}

For the proofs of Lemmas \ref{lem08} and \ref{lem07}, see the Appendix.

\section{Trackability Criteria}\label{sec6}

To develop the basic trackability criteria in ILC, we consider two practical and challenging problems for the system (\ref{b2}) such that it is subject to:
\begin{enumerate}
\item
nonzero system relative degree;

\item
nonzero initial output condition.
\end{enumerate}

\noindent To this end, we notice the physical realization of the system (\ref{b2}) and without loss of generality present the following conditions:
\begin{enumerate}
\item[C1)]
$G_{1}(s)$ and $G_{2}(s)$ are strictly proper;

\item[C2)]
$D(s)=d_{0}+\widehat{D}(s)$ holds for some nonzero vector $d_{0}\in\mathbb{R}^{m}$ and some strictly proper vector $\widehat{D}(s)\in\R\F^{m\times 1}(s)$.
\end{enumerate}

If we denote $\Phi_{1}(t)=\mathcal{L}^{-1}\left[G_{1}(s)\right]$ and $\Phi_{2}(t)=\mathcal{L}^{-1}\left[G_{2}(s)\right]$, then we can arrive at that $\Phi_{1}(t)$ and $\Phi_{2}(t)$ are smooth, namely, $\Phi_{1}^{(j)}(t)$ and $\Phi_{2}^{(j)}(t)$ exist for all $j\in\mathbb{Z}_{+}$ from the condition C1). In the condition C2), $d_{0}$ is closely related with the initial output condition, and we can actually gain $d(t)=d_{0}\delta(t)+\widehat{d}(t)$, where $\widehat{d}(t)\triangleq\mathcal{L}^{-1}\left[\widehat{D}(s)\right]$. Then we can present the following lemma by resorting to the properties of Laplace transform, especially the initial-value theorem \cite{lmt:07}.

\begin{lem}\label{lem01}
For the system (\ref{b2}) with $y_{k}(t)\in\mathbb{R}^{q}$ and $u_{k}(t)\in\mathbb{R}^{p}$, two properties hold under the conditions C1) and C2) below.
\begin{enumerate}
\item
In the series form, $G_{1}(s)$ and $G_{2}(s)$ can be written as
\begin{equation}\label{b4}
G_{1}(s)=\sum_{j=0}^{\infty}\Phi_{1}^{(j)}(0)s^{-(j+1)},
G_{2}(s)=\sum_{j=0}^{\infty}\Phi_{2}^{(j)}(0)s^{-(j+1)}.
\end{equation}

\item
In the time domain, the system (\ref{b2}) can be described as
\begin{equation}\label{b5}
\aligned
y_{k}(t)
&=\int_{0}^{t}\Phi_{1}(t-\tau)u_{k}(\tau)d\tau\\
&~~~+\int_{0}^{t}\Phi_{2}(t-\tau)\widehat{d}(\tau)d\tau+\Phi_{2}(t)d_{0},\quad
\forall k\in\mathbb{Z}_{+}.
\endaligned
\end{equation}
%
\end{enumerate}
\end{lem}

From Lemma \ref{lem01}, it is clear that the system relative degree of (\ref{b2}) is not less than one, and the initial output satisfies $y_{k}(0)=\Phi_{2}(0)d_{0}$, $\forall k\in\mathbb{Z}_{+}$. Similarly to Lemma \ref{lem01}, we can also develop a time-domain trackability result for ILC with Definition \ref{defi2}.

\begin{lem}\label{lem02}
Consider the system (\ref{b2}) under the conditions C1) and C2). Then any specified output trajectory $y_{d}(t)$ is trackable (respectively, realizable) in ILC if and only if there exists some (respectively, a unique) input $u_{d}(t)\in\mathbb{R}^{p}$ such that
\begin{equation}\label{b6}
\aligned
\int_{0}^{t}\Phi_{1}(t-\tau)u_{d}(\tau)d\tau
&=y_{d}(t)-\Phi_{2}(t)d_{0}\\
&~~~-\int_{0}^{t}\Phi_{2}(t-\tau)\widehat{d}(\tau)d\tau,\quad\forall t\in[0,T].
\endaligned
\end{equation}
\end{lem}

From Lemma \ref{lem02}, we note that the trackability of the specified output trajectory in ILC requires the integral equation (\ref{b6}) to be satisfied not only at some instant but also over the whole time interval $[0,T]$. Of particular note is that the trackable trajectory $y_{d}(t)$ should satisfy $y_{d}(0)=\Phi_{2}(0)d_{0}$ for the system (\ref{b2}). Thus, we can conclude from Lemmas \ref{lem01} and \ref{lem02} that for any trackable trajectory $y_{d}(t)$, the following initial condition needs to hold:
\begin{equation}\label{b7}
y_{d}(0)=y_{k}(0)
,\quad\forall k\in\mathbb{Z}_{+}.
\end{equation}

\noindent This represents exactly the class of identical initial conditions, and Lemma \ref{lem02} also provides explanations on why it is required in realizing the perfect tracking tasks of ILC. Otherwise, if (\ref{b7}) does not hold, then $y_{d}(t)$ is not trackable by Lemma \ref{lem02}. Hence, it is obvious from (\ref{b5}) and (\ref{b6}) that the perfect tracking task (\ref{b1}) does not hold for any input sequence $\{u_{k}(t):k\in\mathbb{Z}_{+}\}$, except for the case that the initial shifts can be fully overcome through certain additional control mechanisms (see, e.g., \cite{pm:91} for ILC with impulsive actions).

\subsection{Specific Criteria}

As is well known, the system relative degree condition plays a fundamentally important role in accomplishing tracking tasks of ILC \cite{acm:07}. In fact, it indicates that the fundamental trackability of ILC has an essential relation with the system relative degree condition. To disclose this fact, we focus on the case of relative degree one for controlled systems, which is the relative degree condition most considered for ILC. Specifically, for the system (\ref{b2}), it has a relative degree of one if and only if (see also \cite{so:91})
\begin{enumerate}
\item[C3)]
$G_{1}(s)$ has a relative degree of one.
\end{enumerate}

By following the discussions of, e.g., \cite{ack:93,so:91}, we know that the relative degree condition C3) is characterized by some full rank matrix. This, together with (\ref{b4}), yields the property shown in the following lemma.

\begin{lem}\label{lem03}
Under the condition C1), the condition C3) holds if and only if $\Phi_{1}(0)$ has full rank. Namely, $\Phi_{1}(0)$ has full-row rank when (\ref{b2}) is over-actuated (that is, $q\leq p$); and otherwise, $\Phi_{1}(0)$ has full-column rank when (\ref{b2}) is under-actuated (that is, $q\geq p$).
\end{lem}

For the case $q\geq p$, we explore Lemma \ref{lem03} to develop a further property of $G_{1}(s)$ with the properties of polynomial matrices.

\begin{lem}\label{lem04}
Let $q\geq p$ and the conditions C1) and C3) hold. Then $G^{\tp}_{1}(s)G_{1}(s)\in\R\F^{p\times p}(s)$ is nonsingular.
\end{lem}

Based on Lemma \ref{lem04}, we establish a trackability result of ILC for the system (\ref{b2}) in the under-actuated case with $q\geq p$.

\begin{thm}\label{thm01}
For the system (\ref{b2}) with $q\geq p$, let the conditions C1)--C3) hold. Then any specified trajectory $y_{d}(t)\in C^{1}_{q}[0,T]$ is trackable in ILC if and only if it can satisfy the initial condition (\ref{b7}) and the following frequency-domain algebraic equation:
\begin{equation}\label{b8}
\left\{I-G_1(s)\left[G^{\tp}_{1}(s)G_{1}(s)\right]^{-1}G^{\tp}_{1}(s)\right\}\left[Y_{d}(s)-G_{2}(s)D(s)\right]=0.
\end{equation}

\noindent Further, for any trackable output trajectory $y_{d}(t)$, the algebraic equation (\ref{b3}) has a unique solution in the form of
\begin{equation}\label{b9}
U_{d}(s)=\left[G^{\tp}_{1}(s)G_{1}(s)\right]^{-1}G^{\tp}_{1}(s)\left[Y_{d}(s)-G_{2}(s)D(s)\right]
\end{equation}

\noindent which can fulfill $u_{d}(t)=\mathcal{L}^{-1}\left[U_{d}(s)\right]\in C_{p}[0,T]$.
\end{thm}

\begin{rem}\label{rem01}
Because the number of the output variables to be controlled is not less than that of the input variables, Theorem \ref{thm01} indicates that the trackable output trajectories for the system (\ref{b2}) are given exactly by the solutions for the algebraic equation (\ref{b8}), where they need to particularly satisfy the initial condition (\ref{b7}). Nevertheless, not any output trajectory $y_{d}(t)$ satisfying the initial condition (\ref{b7}) corresponds to the solution of the algebraic equation (\ref{b8}). Besides, the trackability results of Theorem \ref{thm01} can be validated through the frequency-domain methods especially thanks to leveraging properties of polynomial matrices, which however has not been introduced for ILC to our knowledge. A fact worth highlighting for Theorem \ref{thm01} is that for any trackable trajectory $y_{d}(t)\in C^{1}_{q}[0,T]$, the corresponding input $u_{d}(t)$ needs to be continuous such that $u_{d}(t)\in C_{p}[0,T]$.
\end{rem}

For the case $q\leq p$, we note Lemma \ref{lem03} and denote
\begin{equation}\label{b10}
\Phi_{1}(t)
=\begin{bmatrix}\Phi_{1,1}(t) & \Phi_{1,2}(t)\end{bmatrix}~\hbox{with}~
\left\{\aligned
\Phi_{1,1}(t)&\in\R^{q\times q}\\
\Phi_{1,2}(t)&\in\R^{q\times(p-q)}
\endaligned\right.
\end{equation}

\noindent and then, without loss of generality, we give a further condition of the condition C3) as follows:
\begin{enumerate}
\item[C4)]
when $q\leq p$, let $\Phi_{1}(t)$ be given in the block form of (\ref{b10}) such that $\Phi_{1,1}(0)\in\mathbb{R}^{q\times q}$ is nonsingular.
\end{enumerate}

\noindent Otherwise, this can be realized with elementary transformation in column of $\Phi_{1}(0)$, which has no influences on our following analyses and results except for the notations. Correspondingly, by (\ref{b10}), we can write $G_1(s)\in\R\F^{q\times p}(s)$ as
\begin{equation}\label{b11}
G_{1}(s)
=\begin{bmatrix}G_{11}(s) & G_{12}(s)\end{bmatrix}~\hbox{with}~
\left\{\aligned
G_{11}(s)&\in\R\F^{q\times q}(s)\\
G_{12}(s)&\in\R\F^{q\times(p-q)}(s).
\endaligned\right.
\end{equation}

\noindent For (\ref{b11}), we present a nonsingularity property of $G_{11}(s)$ based on the properties of polynomial matrices.

\begin{lem}\label{lem05}
Let $q\leq p$ and the conditions C1), C3), and C4) hold. Then $G_{11}(s)$ is nonsingular.
\end{lem}

With Lemma \ref{lem05}, we propose a trackability result of ILC for the system (\ref{b2}) in the over-actuated case with $q\leq p$.

\begin{thm}\label{thm02}
For the system (\ref{b2}) with $q\leq p$, let the conditions C1)--C4) hold. Then any specified trajectory $y_{d}(t)\in C^{1}_{q}[0,T]$ is trackable in ILC if and only if it can satisfy the initial condition (\ref{b7}). Moreover, the set of the solutions to the algebraic equation (\ref{b3}), i.e., that of the desired inputs for the system (\ref{b2}) to generate the trackable trajectory $y_{d}(t)$, is given by
\begin{equation}\label{b12}
\aligned
\mathcal{U}_{d}
&=\left\{U_{d}(s)\big|G_{1}U_{d}(s)=Y_{d}(s)-G_{2}(s)D(s)\right\}\\
&=\left\{
U_{d}(s)=\begin{bmatrix}G^{-1}_{11}(s)\left[Y_{d}(s)-G_{2}(s)D(s)\right]\\0\end{bmatrix}\right.\\
&~~~~~~~~~~~~~~\left.+\begin{bmatrix}-G^{-1}_{11}(s)G_{12}(s)\\I
\end{bmatrix}U_{d,2}(s)\Big|u_{d,2}(t)\in\R^{p-q}\right\}.
\endaligned
\end{equation}

\noindent In particular, $u_{d}(t)\in C_{p}[0,T]$ if and only if $u_{d,2}(t)\in C_{p-q}[0,T]$.
\end{thm}

\begin{rem}\label{rem02}
In comparison to Theorem \ref{thm01}, Theorem \ref{thm02} presents a quite different trackability result for ILC though they employ the same system relative degree condition. Because the number of the output variables to be controlled is not more than that of the input variables, Theorem \ref{thm02} states that any specified output trajectory fulfilling the initial condition (\ref{b7}) is trackable for the system (\ref{b2}). Namely, any specified output trajectory is trackable in ILC for the initial time $t=0$ if and only if it is trackable in ILC within any time interval $t\in[0,T]$. In accordance with this property, there generally exist multiple inputs that can generate the trackable output trajectory for the system (\ref{b2}). Furthermore, it indicates by (\ref{b12}) that since $q\leq p$, there are $q$ input variables essentially required to achieve the tracking task for any output with $q$ variables, whereas the other $p-q$ input variables can be freely chosen. This actually provides inspiration for the design and analysis of ILC in the presence of over-actuated systems, where how to find input variables that essentially work for the output tracking tasks is crucial.
\end{rem}

Based on Theorems \ref{thm01} and \ref{thm02}, we can also explore the relation between trackability and realizability in ILC as a direct result.

\begin{cor}\label{cor01}
Consider the system (\ref{b2}), and let the conditions C1)--C3) hold. Then for any specified output trajectory $y_{d}(t)\in C^{1}_q[0,T]$ that fulfills the initial condition (\ref{b7}),
\begin{enumerate}
\item
when $q\geq p$, $y_{d}(t)$ is realizable in ILC if and only if it is trackable in ILC; 

\item
when $q<p$, $y_{d}(t)$ is trackable in ILC, but not realizable in ILC.
\end{enumerate}
\end{cor}

\begin{rem}\label{rem03}
From Corollary \ref{cor01}, it follows that trackability and realizability are equivalent in ILC of the under-actuated system (\ref{b2}) under the conditions C1)--C3). However, when the system (\ref{b2}) is over-actuated, realizability no longer makes sense owing to the existence of multiple inputs that can yield any trackable output trajectory. These observations indicate that trackability plays a more fundamental role than realizability in performing the ILC analysis.
\end{rem}

\subsection{State-Space Case Studies}

For the system (\ref{b2}) under the conditions C1)--C3), we assume a time-domain realization in the form of
\begin{equation}\label{a3}
\left\{
\aligned
\dot{x}_{k}(t)
&=Ax_{k}(t)+Bu_{k}(t)+w(t)\\
y_{k}(t)
&=Cx_{k}(t)
\endaligned,\quad\forall t\in[0,T],\forall k\in\mathbb{Z}_{+}
\right.
\end{equation}

\noindent where $x_{k}(t)\in\mathbb{R}^{n}$ is the system state with $x_{k}(0)\triangleq x_{0}$, $\forall k\in\mathbb{Z}_{+}$, $w(t)\in\mathbb{R}^{n}$ is the external disturbance, and $A\in\mathbb{R}^{n\times n}$, $B\in\mathbb{R}^{n\times p}$, and $C\in\mathbb{R}^{q\times n}$ are the system matrices. By the relation between (\ref{b2}) and (\ref{a3}), we know that if let $m=2n$, then
\begin{equation}\label{a4}
\aligned
D(s)&=\left[x_{0}^{\tp}~~W^{\tp}(s)\right]^{\tp}\in\R\F^{2n\times1}(s)\\
G_{1}(s)&=C\left(sI-A\right)^{-1}B\in\R\F^{q \times p}(s)\\
G_{2}(s)&=\left[C\left(sI-A\right)^{-1}~~C\left(sI-A\right)^{-1}\right]\in\R\F^{q\times2n}(s).
\endaligned
\end{equation}

\noindent where $W(s)\triangleq\mathcal{L}\left[w(t)\right]$. We can also obtain from (\ref{a4}) that
\[
\Phi_{1}(t)=Ce^{At}B,\quad
\Phi_{2}(t)=\left[Ce^{At}~~Ce^{At}\right].
\]

\noindent Of note is that (\ref{a3}) is one of the commonly considered systems in continuous-time ILC (for more details, see the survey \cite{acm:07}).

Thanks to (\ref{a4}), we have for the system (\ref{a3}) that
\begin{enumerate}
\item[i)]
C1) naturally holds;

\item[ii)]
C2) holds if and only if we set $d_{0}=\left[x_{0}^{\tp}~~0\right]^{\tp}$ and $\widehat{D}(s)=\left[0~~W^{\tp}(s)\right]^{\tp}$;

\item[iii)]
C3) holds if and only if $CB$ has full rank.
%
\end{enumerate}

\noindent In addition, for $q\leq p$, if we denote $B=\left[B_{1}~B_{2}\right]$ with $B_{1}\in\R^{n\times q}$ and $B_{2}\in\R^{n\times(p-q)}$, then
\begin{enumerate}
\item[iv)]
C4) holds if and only if $CB_{1}$ is nonsingular.
\end{enumerate}

\noindent With the properties i) and ii), we know from Lemma \ref{lem02} that for the system (\ref{a3}), any specified output trajectory $y_{d}(t)\in C^{1}_q[0,T]$ is trackable in ILC if and only if there exists some $u_{d}(t)$ such that, for all $t\in[0,T]$,
\begin{equation}\label{a7}
\aligned
\int_{0}^{t}Ce^{A(t-\tau)}Bu_{d}(\tau)d\tau
&=y_{d}(t)-Ce^{At}x_{0}\\
&~~~-\int_{0}^{t}Ce^{A(t-\tau)}w(\tau)d\tau.
\endaligned
\end{equation}

\noindent Clearly, the initial condition (\ref{b7}) becomes $y_{d}(0)=y_{k}(0)=Cx_{0}$, $\forall k\in\mathbb{Z}_{+}$, which also coincides with the facts of (\ref{a3}) and (\ref{a7}). In addition, we can directly establish the following trackability result of ILC as a consequence of Theorems \ref{thm01} and \ref{thm02}.

\begin{cor}\label{cor02}
For the system (\ref{a3}), the following results hold for any specified output trajectory $y_{d}(t)\in C^{1}_{q}[0,T]$.
\begin{enumerate}
\item
When $q\geq p$, let $CB$ be of full-column rank. Then $y_{d}(t)$ is trackable in ILC if and only if $y_{d}(0)=Cx_{0}$ holds and its Laplace transform $Y_{d}(s)=\mathcal{L}[y_{d}(t)]$ fulfills the algebraic equation (\ref{b8}), where $D(s)$, $G_{1}(s)$ and $G_{2}(s)$ are defined by (\ref{a4}). Further, there exists a unique input correspondingly generate any trackable output trajectory.

\item
When $q\leq p$, let $CB$ be of full-row rank. Then there exist multiple inputs such that $y_{d}(t)$ is trackable in ILC if and only if it satisfies $y_{d}(0)=Cx_{0}$.
\end{enumerate}
\end{cor}

In Corollary \ref{cor02}, we reveal that the trackability is tied closely with the full rank of $CB$ for continuous-time linear ILC in the presence of the relative degree one. It actually provides a basic guarantee for the existing ILC design results (for more details, see technical overview of ILC in \cite{acm:07}), and a clear explanation on why they work effectively in realizing the tracking tasks.

\subsection{Technical Proofs}

Next, we give detailed proofs of Lemmas \ref{lem01}--\ref{lem05} and Theorems \ref{thm01} and \ref{thm02}, especially by resorting to a frequency-domain analysis method with properties of polynomial matrices.

\begin{IEEEproof}[Proof of Lemma \ref{lem01}]
By the condition C1), we can denote $G_{1}(s)$ and $G_{2}(s)$ in the series form of
\begin{equation*}\label{}
G_{1}(s)=\sum_{j=0}^{\infty}g_{1,j}s^{-(j+1)},\quad G_{2}(s)=\sum_{j=0}^{\infty}g_{2,j}s^{-(j+1)}
\end{equation*}

\noindent for some matrices $g_{1,j}\in\mathbb{R}^{q\times p}$ and $g_{2,j}\in\mathbb{R}^{q\times m}$, $\forall j\in\mathbb{Z}_{+}$. Thus, taking the Laplace transformation leads to
\begin{equation*}\label{}
\Phi_{1}(t)=\sum_{j=0}^{\infty}g_{1,j}\frac{\Ds t^{j}}{\Ds j!},\quad \Phi_{2}(t)=\sum_{j=0}^{\infty}g_{2,j}\frac{\Ds t^{j}}{\Ds j!}
\end{equation*}

\noindent from which we can easily validate
\begin{equation*}\label{}
g_{1,j}=\Phi_{1}^{(j)}(0),\quad g_{2,j}=\Phi_{2}^{(j)}(0),\quad\forall j\in\mathbb{Z}_{+}
\end{equation*}

\noindent and consequently it is immediate to gain (\ref{b4}). By the conditions C1) and C2), (\ref{b5}) can be equivalently developed from (\ref{b2}) thanks to taking the inverse Laplace transformation.
\end{IEEEproof}

\begin{IEEEproof}[Proof of Lemma \ref{lem02}]
A direct result of Definition \ref{defi2}.
\end{IEEEproof}

\begin{IEEEproof}[Proof of Lemma \ref{lem03}]
Due to (\ref{b4}) in Lemma \ref{lem01}, this lemma is a consequence of the condition C3) according to the definition of the system relative degree (see also \cite{ack:93,so:91}).
\end{IEEEproof}

\begin{IEEEproof}[Proof of Lemma \ref{lem04}]
Under the condition C1), we consider a realization of $G_{1}(s)$, which without any loss of generality is denoted by $\left(A\in\R^{n\times n},B\in\R^{n\times p},C\in\R^{q\times n}\right)$. Namely, we have $G_{1}(s)=C\left(sI-A\right)^{-1}B$, with which we can validate $\Phi_{1}^{(j)}(0)=CA^{j}B$, $\forall j\in\mathbb{Z}_{+}$. Thus, a consequence of (\ref{b4}) is such that
\begin{equation}\label{a26}
\aligned
s^{i}G_{1}(s)
&=\sum_{j=0}^{i-1}CA^{j}Bs^{i-1-j}\\
&~~~+C(sI-A)^{-1}A^{i}B,\quad\forall i=0,1,\cdots,n.
\endaligned
\end{equation}

\noindent Let the characteristic polynomial of $A$ be $\alpha(s)=\det\left(sI-A\right)\triangleq\sum_{i=0}^{n}\alpha_{i}s^{i}$ for some $\alpha_{i}\in\mathbb{R}$, $\forall i=0$, $1$, $\cdots$, $n-1$ and $\alpha_{n}=1$, and then with the Cayley-Hamilton theorem, we have $\sum_{i=0}^{n}\alpha_{i}A^{i}=0$ which, together with (\ref{a26}), leads to
\begin{equation}\label{a27}
\aligned
\sum_{i=0}^{n}\alpha_{i}s^{i}G_{1}(s)
&=\sum_{i=0}^{n}\alpha_{i}\sum_{j=0}^{i-1}CA^{j}Bs^{i-1-j}
+\sum_{i=0}^{n}\alpha_{i}C(sI-A)^{-1}A^{i}B\\
&=\sum_{i=0}^{n}\alpha_{i}\sum_{j=0}^{i-1}\Phi_{1}^{(j)}(0)s^{i-1-j}\\
&\triangleq Q(s)
\endaligned
\end{equation}

\noindent where $Q(s)\in \R\P^{q \times p}(s)$ (see \cite[p. 524]{am:06}) satisfies
\begin{equation}\label{a28}
\aligned
Q(s)&=\sum_{i=0}^{n}\alpha_{i}\sum_{j=0}^{i-1}\Phi_{1}^{(j)}(0)s^{i-1-j}\\
&=\alpha_{n}\Phi_{1}(0)s^{n-1}+\alpha_{n}\sum_{j=1}^{n-1}\Phi_{1}^{(j)}(0)s^{n-1-j}\\
&~~~+\sum_{i=0}^{n-1}\alpha_{i}\sum_{j=0}^{i-1}\Phi_{1}^{(j)}(0)s^{i-1-j}\\
&=\alpha_{n}\Phi_{1}(0)s^{n-1}+\sum_{i=0}^{n-2}\left[\sum_{j=i+1}^{n}\alpha_{j}\Phi_{1}^{(j-i-1)}(0)\right]s^{i}\\
&=\Phi_{1}(0)s^{n-1}+\sum_{i=0}^{n-2}\left[\sum_{j=i+1}^{n}\alpha_{j}\Phi_{1}^{(j-i-1)}(0)\right]s^{i}.
\endaligned
\end{equation}

\noindent Clearly, $\Phi_{1}(0)$ is the highest column degree coefficient matrix of $Q(s)$ (see \cite[p. 526]{am:06}). Owing to $q\geq p$, we can obtain from Lemma \ref{lem03} that $\Phi_{1}(0)$ has full-column rank under the condition C3), and hence $Q(s)$ is column reduced or column proper (see \cite[p. 527]{am:06}). Then by (\ref{a28}), $Q^{\tp}(s)Q(s)$ is nonsingular because we can use the result (2.4) of \cite[p. 527]{am:06} to arrive at
\[\aligned
\det\left(Q^{\tp}(s)Q(s)\right)
&=\det\left(\Phi_{1}^{\tp}(0)\Phi_{1}(0)\right)s^{(2n-2)p}\\
&~~~+\hbox{lower degree terms}\\
&\not\equiv0.
\endaligned\]

\noindent This immediately leads to that $G^{\tp}_{1}(s)G_{1}(s)$ is nonsingular since the use of (\ref{a27}) results in
\[
G^{\tp}_{1}(s)G_{1}(s)=\left(\sum_{i=0}^{n}\alpha_{i}s^{i}\right)^{-2}Q^{\tp}(s)Q(s).
\]

\noindent That is, the proof of Lemma \ref{lem04} is completed.
\end{IEEEproof}

\begin{IEEEproof}[Proof of Lemma \ref{lem05}]
Let $Q(s)\triangleq\left[Q_{1}(s)~~Q_{2}(s)\right]$ for $Q_{1}(s)\in\R\P^{q\times q}(s)$ and $Q_{2}(s)\in\R\P^{q\times(p-q)}(s)$. Then from (\ref{b11}) and (\ref{a27}), we have
\begin{equation}\label{a29}
\aligned
Q_{h}(s)
&=\sum_{i=0}^{n}\alpha_{i}s^{i}G_{1h}(s)\\
&=\sum_{i=0}^{n}\alpha_{i}\sum_{j=0}^{i-1}\Phi_{1,h}^{(j)}(0)s^{i-1-j},\quad\forall h\in\{1,2\}
\endaligned
\end{equation}

\noindent with which we follow the same lines as (\ref{a28}) to further get
\begin{equation}\label{a30}
\aligned
Q_{h}(s)
&=\Phi_{1,h}(0)s^{n-1}\\
&~~~+\sum_{i=0}^{n-2}\left(\sum_{j=i+1}^{n}\alpha_{j}\Phi_{1,h}^{(j-i-1)}(0)\right)s^{i},\quad\forall h\in\{1,2\}.
\endaligned
\end{equation}

\noindent Thanks to considering the result (2.4) of \cite[p. 527]{am:06} for $Q_{1}(s)$, we can leverage (\ref{a30}) and adopt the condition C4) to arrive at
\[\aligned
\det\left(Q_{1}(s)\right)
&=\det\left(\Phi_{1,1}(0)\right)s^{(n-1)q}+\hbox{lower degree terms}\\
&\not\equiv0
\endaligned\]

\noindent which ensures that $Q_{1}(s)$ is nonsingular. Note also that the use of (\ref{a29}) leads to
\begin{equation}\label{a31}
G_{1h}(s)=\left(\sum_{i=0}^{n}\alpha_{i}s^{i}\right)^{-1}Q_{h}(s),\quad\forall h\in\{1,2\}.
\end{equation}

\noindent As a direct consequence of (\ref{a31}), $G_{11}(s)$ is nonsingular.
\end{IEEEproof}

\begin{IEEEproof}[Proof of Theorem \ref{thm01}]
We first prove the equivalent relation between the trackability of a specified output trajectory $y_{d}(t)\in C^{1}_{q}[0,T]$ and the satisfactions of the initial condition (\ref{b7}) and the frequency-domain algebraic equation (\ref{b8}) by $Y_{d}(s)=\mathcal{L}\left[y_{d}(t)\right]$.

{\it Necessity:} With Definition \ref{defi2}, the trackability of the specified trajectory $y_{d}(t)$ implies that there exists some input $u_{d}(t)\in\mathbb{R}^{p}$ for the system (\ref{b2}) to guarantee the satisfaction of the algebraic equation (\ref{b3}) by $U_{d}(s)=\mathcal{L}\left[u_{d}(t)\right]$. Then thanks to Lemma \ref{lem04}, we can multiply $I-G_1(s)\left[G^{\tp}_{1}(s)G_{1}(s)\right]^{-1}G^{\tp}_{1}(s)$ on both sides of (\ref{b3}) such that
\[
\aligned
&\left\{I-G_1(s)\left[G^{\tp}_{1}(s)G_{1}(s)\right]^{-1}G^{\tp}_{1}(s)\right\}\left[Y_{d}(s)-G_{2}(s)D(s)\right]\\
&~~~=\left\{I-G_1(s)\left[G^{\tp}_{1}(s)G_{1}(s)\right]^{-1}G^{\tp}_{1}(s)\right\}G_{1}(s)U_{d}(s)\\
&~~~=0
\endaligned
\]

\noindent i.e., the algebraic equation (\ref{b8}) is satisfied by $Y_{d}(s)=\mathcal{L}\left[y_{d}(t)\right]$. In addition, by (\ref{b6}) in Lemma \ref{lem02}, $y_{d}(0)=\Phi_{2}(0)d_{0}$ is immediate, and due to (\ref{b5}) in Lemma \ref{lem01}, we have $y_{k}(0)=\Phi_{2}(0)d_{0}$, $\forall k\in\mathbb{Z}_{+}$. This ensures that $y_{d}(t)$ fulfills the initial condition (\ref{b7}).

{\it Sufficiency:} From Lemma \ref{lem04}, it is feasible that for the system (\ref{b2}) under the conditions C1)--C3), $Y_{d}(s)=\mathcal{L}\left[y_{d}(t)\right]$ fulfills the algebraic equation (\ref{b8}) for the specified trajectory $y_{d}(t)$. Hence, we use (\ref{b8}) to equivalently arrive at
\[
\aligned
G_1(s)\left[G^{\tp}_{1}(s)G_{1}(s)\right]^{-1}G^{\tp}_{1}(s)&\left[Y_{d}(s)-G_{2}(s)D(s)\right]\\
&=Y_{d}(s)-G_{2}(s)D(s)
\endaligned
\]

\noindent from which taking $U_d(s)$ in (\ref{b9}) immediately leads to
\[\aligned
G_1(s)U_d(s)
&=G_1(s)(G^{\tp}_{1}(s)G_{1}(s))^{-1}G^{\tp}_{1}(s)\left[Y_{d}(s)-G_{2}(s)D(s)\right]\\
&=Y_{d}(s)-G_{2}(s)D(s).
\endaligned\]

\noindent That is, $U_d(s)$ is a solution for the algebraic equation (\ref{b3}). From (\ref{b5}) in Lemma \ref{lem01}, we can gain $y_{k}(0)=\Phi_{2}(0)d_{0}$, $\forall k\in\mathbb{Z}_{+}$, which together with the initial condition (\ref{b7}) leads to $y_{d}(0)=\Phi_{2}(0)d_{0}$. We can consequently obtain
\begin{equation}\label{a49}
sY_{d}(s)
=\mathcal{L}\left[\dot{y}_{d}(t)\right]+y_{d}(0)
=\mathcal{L}\left[\dot{y}_{d}(t)\right]+\Phi_{2}(0)d_{0}.
\end{equation}

\noindent Since we can rewrite $U_d(s)$ in (\ref{b9}) as
\[\aligned
U_{d}(s)&=s^{-1}\left[G^{\tp}_{1}(s)G_{1}(s)\right]^{-1}G^{\tp}_{1}(s)\left[sY_{d}(s)\right]\\
&~~~-\left[G^{\tp}_{1}(s)G_{1}(s)\right]^{-1}G^{\tp}_{1}(s)G_{2}(s)D(s)
\endaligned\]

\noindent then under the condition C2) and with (\ref{a49}), we can deduce
\begin{equation}\label{a50}
\aligned
U_{d}(s)
&=s^{-1}\left[G^{\tp}_{1}(s)G_{1}(s)\right]^{-1}G^{\tp}_{1}(s)\mathcal{L}\left[\dot{y}_{d}(t)\right]\\
&~~~-\left[G^{\tp}_{1}(s)G_{1}(s)\right]^{-1}G^{\tp}_{1}(s)G_{2}(s)\widehat{D}(s)\\
&~~~+s^{-1}\left[G^{\tp}_{1}(s)G_{1}(s)\right]^{-1}G^{\tp}_{1}(s)\left[\Phi_{2}(0)-sG_{2}(s)\right]d_{0}.
\endaligned
\end{equation}

\noindent When $q\geq p$, we can obtain from Lemma \ref{lem03} that $\Phi_{1}^{\tp}(0)\Phi_{1}(0)$ is nonsingular under the conditions C1) and C3). Then by noting (\ref{a27}) and (\ref{a28}), we can verify
\begin{equation}
\label{a51}
\aligned
\lim_{s\to\infty}&s^{-1}\left[G^{\tp}_{1}(s)G_{1}(s)\right]^{-1}G^{\tp}_{1}(s)\\
&=\lim_{s\to\infty}\left\{\left[sG_{1}(s)\right]^{\tp}\left[sG_{1}(s)\right]\right\}^{-1}\left[sG^{\tp}_{1}(s)\right]\\
&=\left[\Phi_{1}^{\tp}(0)\Phi_{1}(0)\right]^{-1}\Phi_{1}^{\tp}(0).
\endaligned
\end{equation}

\noindent This, together with the conditions C1) and C2), leads to
\begin{equation}\label{a52}
\aligned
\lim_{s\to\infty}&\left[G^{\tp}_{1}(s)G_{1}(s)\right]^{-1}G^{\tp}_{1}(s)G_{2}(s)\widehat{D}(s)\\
&=\lim_{s\to\infty}\left\{s^{-1}\left[G^{\tp}_{1}(s)G_{1}(s)\right]^{-1}G^{\tp}_{1}(s)\right\}\lim_{s\to\infty}\left[sG_{2}(s)\right]\lim_{s\to\infty}\widehat{D}(s)\\
&=0
\endaligned
\end{equation}

\noindent where $\lim_{s\to\infty}\left[sG_{2}(s)\right]=\Phi_{2}(0)$ due to (\ref{b4}) and $\lim_{s\to\infty}\widehat{D}(s)=0$ are incorporated. In a similar way as (\ref{a52}), we can derive
\begin{equation}\label{a5}
\lim_{s\to\infty}s^{-1}\left[G^{\tp}_{1}(s)G_{1}(s)\right]^{-1}G^{\tp}_{1}(s)\left[\Phi_{2}(0)-sG_{2}(s)\right]=0.
\end{equation}

\noindent By incorporating (\ref{a51})--(\ref{a5}) into (\ref{a50}), we can employ the result of Lemma \ref{lem07} to get $u_{d}(t)=\mathcal{L}^{-1}\left[U_{d}(s)\right]\in C_{p}[0,T]$. Thus, $U_d(s)$ in (\ref{b9}) not only fulfills (\ref{b3}) but also yields $u_{d}(t)=\mathcal{L}^{-1}\left[U_{d}(s)\right]\in\mathbb{R}^{p}$. Then in view of Definition \ref{defi2}, the specified trajectory $y_{d}(t)$ is trackable in ILC for the system (\ref{b2}).

Next, we adopt a proof by contradiction to show that if $y_{d}(t)$ is trackable, then the solution for the algebraic equation (\ref{b3}) is unique. Hence, in addition to $u_{d}(t)$ obtained by (\ref{b9}), we assume an input $\widehat{u}_{d}(t)\in\mathbb{R}^{p}$ different from $u_{d}(t)$ (that is, $\widehat{u}_{d}(t)\neq u_{d}(t)$) such that $\widehat{U}_{d}(s)=\mathcal{L}\left[\widehat{u}_{d}(t)\right]$ also satisfies (\ref{b3}), namely,
\begin{equation}\label{a53}
Y_{d}(s)-G_{2}(s)D(s)=G_{1}(s)\widehat{U}_{d}(s).
\end{equation}

\noindent Then the use of (\ref{b3}) and (\ref{a53}) yields
\[
\aligned
G^{\tp}_{1}(s)G_{1}(s)U_{d}(s)&=G^{\tp}_{1}(s)[Y_{d}(s)-G_{2}(s)D(s)]\\
&=G^{\tp}_{1}(s)G_{1}(s)\hat{U}_{d}(s).
\endaligned
\]

\noindent Since $G^{\tp}_{1}(s)G_{1}(s)$ is nonsingular, we can deduce $U_{d}(s)=\widehat{U}_{d}(s)$ which contradicts the made hypothesis $\widehat{u}_{d}(t)\neq u_{d}(t)$. Thus, we can conversely conclude that the algebraic equation (\ref{b3}) has a unique solution given by (\ref{b9}).
\end{IEEEproof}

\begin{IEEEproof}[Proof of Theorem \ref{thm02}]
{\it Necessity:} With the conditions C1)--C4), if $y_{d}(t)$ is a trackable trajectory in ILC, then from Lemma \ref{lem02}, $y_{d}(0)=\Phi_{2}(0)d_{0}$ holds as a consequence of (\ref{b6}). Because (\ref{b5}) yields $y_{k}(0)=\Phi_{2}(0)d_{0}$, $\forall k\in\mathbb{Z}_{+}$ by Lemma \ref{lem01}, we immediately know that the initial condition (\ref{b7}) holds.

{\it Sufficiency:} If the initial condition (\ref{b7}) holds, then we employ (\ref{b5}), and can actually obtain $y_{d}(0)=y_{k}(0)=\Phi_{2}(0)d_{0}$, $\forall k\in\mathbb{Z}_{+}$. In view of this result, we will show that the algebraic equation (\ref{b3}) is solvable. Due to $q\leq p$ and with (\ref{b11}), we correspondingly denote $u_{d}(t)$ as $u_{d}(t)=\left[u_{d,1}^{\tp}(t)~~u_{d,2}^{\tp}(t)\right]^{\tp}$ for $u_{d,1}(t)\in\mathbb{R}^{q}$ and $u_{d,2}(t)\in\mathbb{R}^{p-q}$. Let us also denote $U_{d}(s)=\left[U_{d,1}^{\tp}(s)~~U_{d,2}^{\tp}(s)\right]^{\tp}$, where $U_{d,1}(s)\triangleq\mathcal{L}\left[u_{d,1}(t)\right]$ and $U_{d,2}(s)=\mathcal{L}\left[u_{d,2}(t)\right]$. Hence, by incorporating (\ref{b11}), we can equivalently derive from (\ref{b3}) that
\begin{equation}\label{a37}
G_{11}(s)U_{d,1}(s)=Y_{d}(s)-G_{2}(s)D(s)-G_{12}(s)U_{d,2}(s).
\end{equation}

\noindent Because $G_{11}(s)$ is nonsingular from Lemma \ref{lem05}, we can leverage (\ref{a37}) to arrive at
\begin{equation}
\label{a38}
U_{d,1}(s)
=G_{11}^{-1}(s)\left[Y_{d}(s)-G_{2}(s)D(s)-G_{12}(s)U_{d,2}(s)\right]
\end{equation}

\noindent which straightforwardly results in
\[
U_{d}(s)
=\begin{bmatrix}
     G^{-1}_{11}(s)[Y_{d}(s)-G_{2}(s)D(s)-G_{12}(s)U_{d,2}(s)]\\
    U_{d,2}(s)
  \end{bmatrix}.
\]

\noindent Consequently, $y_{d}(t)$ is trackable in ILC, and we can determine the solutions for the algebraic equation (\ref{b3}) with (\ref{b12}) by taking any $u_{d,2}(t)\in\mathbb{R}^{p-q}$.

Next, we prove that $u_{d}(t)\in C_{p}[0,T]$ if $u_{d,2}(t)\in C_{p-q}[0,T]$. By inserting (\ref{a49}), we can rewrite (\ref{a38}) as
\begin{equation}\label{a6}
\aligned
U_{d,1}(s)
&=\left[sG_{11}(s)\right]^{-1}\left[sY_{d}(s)\right]-G_{11}^{-1}(s)G_{2}(s)D(s)\\
&~~~-G_{11}^{-1}(s)G_{12}(s)U_{d,2}(s)\\
&=\left[sG_{11}(s)\right]^{-1}\mathcal{L}\left[\dot{y}_{d}(t)\right]
-G_{11}^{-1}(s)G_{2}(s)\widehat{D}(s)\\
&~~~+\left[sG_{11}(s)\right]^{-1}\left[\Phi_{2}(0)-sG_{2}(s)\right]d_{0}\\
&~~~-G_{11}^{-1}(s)G_{12}(s)U_{d,2}(s).
\endaligned
\end{equation}

\noindent From (\ref{b4}) and (\ref{b11}), it is clear to see $\lim_{s\to\infty}\left[sG_{11}(s)\right]=\Phi_{1,1}(0)$, $\lim_{s\to\infty}\left[sG_{12}(s)\right]=\Phi_{1,2}(0)$ and $\lim_{s\to\infty}\left[sG_{2}(s)\right]=\Phi_{2}(0)$. Then under the condition C4), we have
\begin{equation}\label{a39}
\aligned
\lim_{s\to\infty}\left[sG_{11}(s)\right]^{-1}=\Phi_{1,1}^{-1}(0)
\endaligned
\end{equation}

\noindent which further leads to
\begin{equation}\label{a40}
\aligned
&\lim_{s\to\infty}\left[sG_{11}(s)\right]^{-1}\left[\Phi_{2}(0)-sG_{2}(s)\right]=0\\
&\lim_{s\to\infty}G_{11}^{-1}(s)G_{12}(s)=\Phi_{1,1}^{-1}(0)\Phi_{1,2}(0).
\endaligned
\end{equation}

\noindent For the same reason as (\ref{a40}), we can employ the condition C2) to arrive at
\begin{equation}\label{a41}
\aligned
\lim_{s\to\infty}G_{11}^{-1}(s)G_{2}(s)\widehat{D}(s)
&=\lim_{s\to\infty}\left[sG_{11}(s)\right]^{-1}\lim_{s\to\infty}\left[sG_{2}(s)\right]\lim_{s\to\infty}\widehat{D}(s)\\
&=0.
\endaligned
\end{equation}

\noindent By incorporating (\ref{a39})--(\ref{a41}) into (\ref{a6}), we benefit from Lemma \ref{lem07} to obtain that $u_{d,1}(t)=\mathcal{L}^{-1}\left[U_{d,1}(s)\right]\in C_{q}[0,T]$ if $u_{d,2}(t)=\mathcal{L}^{-1}\left[U_{d,2}(s)\right]\in C_{p-q}[0,T]$. As a consequence, it follows that $u_{d}(t)\in C_{p}[0,T]$ if and only if $u_{d,2}(t)\in C_{p-q}[0,T]$.
\end{IEEEproof}

\section{Trackability-Based ILC Synthesis}\label{sec3}

In this section, we first introduce an ILC updating law with a feedback-based design method and then explore the developed trackability results to implement the corresponding ILC design and analysis. In particular, we utilize an FCS-induced approach to establish the ILC convergence analysis from the viewpoints of both output and input through a unified condition, regardless of under-actuated or over-actuated MIMO controlled systems.

\subsection{Trackability-Based ILC Results}

For the tracking task (\ref{b1}), let the tracking error be represented as $e_{k}(t)=y_{d}(t)-y_{k}(t)$. Then it clearly becomes $\lim_{k\to\infty}e_{k}(t)=0$, $\forall t\in[0,T]$, which can actually be seen as a class of ``$k$-state stability'' problems arising from the tracking tasks for ILC (see also \cite{mw:21} for similar discussions). Let $E_{k}(s)=\mathcal{L}\left[e_{k}(t)\right]$, which obviously satisfies $E_{k}(s)=Y_{d}(s)-Y_{k}(s)$, and consequently, the iteration-domain dynamics of it can be described by
\[
E_{k+1}(s)=E_{k}(s)-\Delta Y_{k}(s),\quad\forall k\in\mathbb{Z}_{+}.
\]

\noindent If we consider the system (\ref{b2}), then we further have
\[
E_{k+1}(s)=E_{k}(s)-G_{1}(s)\Delta U_{k}(s),\quad\forall k\in\mathbb{Z}_{+}
\]

\noindent where $\Delta U_{k}(s)$ plays the rose as an input to stabilize the $k$-state $E_{k}(s)$ from the viewpoint of iteration-domain dynamics. Thus, we employ the feedback-based design theory and can propose
\[
\Delta U_{k}(s)=\Gamma(s)E_{k}(s), \quad\forall k\in\mathbb{Z}_{+}
\]

\noindent which equivalently yields an ILC updating law in the form of
\begin{equation}\label{a12}
U_{k+1}(s)=U_{k}(s)+\Gamma(s)E_{k}(s), \quad\forall k\in\mathbb{Z}_{+}
\end{equation}

\noindent with $\Gamma(s)\in\R\F^{p\times q}(s)$ as a gain matrix operator to be designed.

\begin{rem}\label{rem04}
In continuous-time linear ILC, we accomplish the design of updating laws by incorporating the general feedback-based design method. This bridges an explicit relation between the design methods of ILC and classic feedback-based control. In particular, (\ref{a12}) involves classic PID-type ILC updating laws as special cases. For example, taking $\Gamma(s)=s\Upsilon$ leads to the D-type ILC updating law from (\ref{a12}), where $\Upsilon\in\R^{p\times q}$ is constant.
\end{rem}

To proceed, we consider applying the ILC updating law (\ref{a12}) to the system (\ref{b2}) and, consequently, can arrive at some design conditions of the gain matrix operator $\Gamma(s)$.

\begin{lem}\label{lem06}
For the system (\ref{b2}) under the conditions C1)--C3), let the ILC updating law (\ref{a12}) be applied under any initial input $U_{0}(s)=\mathcal{L}\left[u_{0}(t)\right]$ for $u_{0}(t)\in C_{p}[0,T]$ and any specified output trajectory $y_{d}(t)\in C^{1}_{q}[0,T]$. Then the following three conditions are equivalent:
\begin{enumerate}
\item
$\Gamma(s)$ is such that $\Gamma(s)G_{1}(s)$ is proper;

\item
$\Gamma(s)$ is such that $G_{1}(s)\Gamma(s)$ is proper;

\item
$s^{-1}\Gamma(s)$ is proper, that is, $s^{-1}\Gamma(s)=\Gamma_{0}+\widehat{\Gamma}(s)$ holds for some nonzero matrix $\Gamma_{0}\in\mathbb{R}^{p\times q}$ and some strictly proper matrix $\widehat{\Gamma}(s)\in\mathbb{RF}^{p\times q}(s)$, where
\begin{equation}\label{b16}
\lim_{s\to\infty}s^{-1}\Gamma(s)=\Gamma_{0}.
\end{equation}
\end{enumerate}

\noindent Further, if any of the abovementioned conditions 1)--3) holds, then in the time domain, (\ref{a12}) reads as
\begin{equation}\label{b17}
\aligned
u_{k+1}(t)
&=u_{k}(t)+\Gamma_{0}\dot{e}_{k}(t)+\int_{0}^{t}\Phi_{\widehat{\Gamma}}(t-\tau)\dot{e}_{k}(\tau)d\tau\\
&~~~+\Phi_{\widehat{\Gamma}}(t)e_{k}(0)+\Gamma_{0}e_{k}(0)\delta(t),\quad\forall k\in\mathbb{Z}_{+},t\in[0,T]
\endaligned
\end{equation}

\noindent where $\Phi_{\widehat{\Gamma}}(t)\triangleq\mathcal{L}^{-1}\left[\widehat{\Gamma}(s)\right]$ and, in particular, it follows that
\begin{equation}\label{b18}
u_{k}(t)\in C_p[0,T],~\forall k\in\mathbb{Z}_{+}\quad\Leftrightarrow\quad\Gamma_{0}e_{k}(0)=0,~\forall k\in\mathbb{Z}_{+}.
\end{equation}
\end{lem}

\begin{rem}\label{rem05}
In Lemma \ref{lem06}, it discloses that under certain design condition, the time-domain realization of the ILC updating law (\ref{a12}) may involve an impulsive mechanism, as revealed by (\ref{b17}). Even though it may help to overcome the initial shift problems for ILC (see, e.g., \cite{pm:91}), the use of the impulsive mechanism is not admissible in practice, as noted in \cite{sy:13}, where it may yield $u_{k}(0)\not\in\mathbb{R}^{p}$ that is not consistent with the condition $u_{d}(t)\in\mathbb{R}^{p}$ needed by the trackable $y_{d}(t)\in\mathbb{R}^{q}$ in Definition \ref{defi2}. Fortunately, the impulsive mechanism resulted in (\ref{b17}) is related only to the initial tracking error that disappears under the initial condition (\ref{b7}). By Theorems \ref{thm01} and \ref{thm02}, it follows that any trackable output trajectory $y_{d}(t)$ for the system (\ref{b2}) satisfies the initial condition (\ref{b7}). This, together with Lemma \ref{lem06}, indicates that a sequence of continuous inputs for ILC can thus be generated to accomplish the tracking tasks under the ILC updating law (\ref{a12}), as revealed by the equivalent relation (\ref{b18}).
\end{rem}

Next, we benefit from Lemma \ref{lem06} to further gain convergence analysis results of ILC with the established trackability criteria. From the perspective of input, applying the ILC updating law (\ref{a12}) to the system (\ref{b2}) leads to
\begin{equation}\label{a16}
\aligned
U_{k+1}(s)
&=\left[I-\Gamma(s)G_{1}(s)\right]U_{k}(s)\\
&~~~+\Gamma(s)\left[Y_{d}(s)-G_{2}(s)D(s)\right],\quad\forall k\in\mathbb{Z}_{+}
\endaligned
\end{equation}

\noindent but, by contrast, from the perspective of output (or equivalently the tracking error), it results in
\begin{equation}\label{b19}
E_{k+1}(s)
=\left[I-G_{1}(s)\Gamma(s)\right]E_{k}(s),\quad\forall k\in\mathbb{Z}_{+}.
\end{equation}

\noindent By the comparison between (\ref{a16}) and (\ref{b19}), different conditions are actually required for the convergence analysis of ILC if it is established from the different perspectives of input and output. In particular, when $q\neq p$, convergence conditions required for the input of (\ref{a16}) and the tracking error of (\ref{b19}) even contradict with each other. Despite this issue, we try to leverage an FCS-induced approach of ILC to arrive at a unified design condition for $\Gamma(s)$ such that we can accomplish the convergence for both input and tracking error, regardless of under-actuated or over-actuated MIMO systems.

Let us revisit (\ref{a16}), and then we can arrive at
\begin{equation}\label{a17}
\aligned
U_{k+2}(s)-U_{k+1}(s)
&=\left[I-\Gamma(s)G_{1}(s)\right]\\
&~~~\times\left[U_{k+1}(s)-U_{k}(s)\right],\quad\forall k\in\mathbb{Z}_{+}.
\endaligned
\end{equation}

\noindent By this development of the input sequence $\{u_{k}(t):k\in\Z_{+}\}$ and based on Lemma \ref{lem08}, we can leverage an FCS-induced approach to present an ILC convergence result in the under-actuated case of the system (\ref{b2}) with $q\geq p$ by employing the ILC trackability result of Theorem \ref{thm01}, as well as the design result of Lemma \ref{lem06}.

\begin{thm}\label{thm03}
For the system (\ref{b2}) with $q\geq p$, let the conditions C1)--C3) be satisfied, and the ILC updating law (\ref{a12}) be applied with any initial input $U_{0}(s)=\mathcal{L}\left[u_{0}(t)\right]$ for $u_{0}(t)\in C_{p}[0,T]$ and any specified output trajectory $y_{d}(t)\in C^{1}_{q}[0,T]$ that satisfies the initial condition (\ref{b7}). Then $u_{k}(t)\in C_p[0,T]$, $\forall k\in\mathbb{Z}_{+}$ is such that $\lim_{k\to\infty}u_{k}(t)=u_{\infty}(t)$ holds for some $u_{\infty}(t)\in C_{p}[0,T]$, together with giving $\lim_{k\to\infty}y_{k}(t)=y_{\infty}(t)$ for some $y_{\infty}(t)\in C^{1}_{q}[0,T]$ as
\[
y_{\infty}(t)
=\int_{0}^{t}\Phi_{1}(t-\tau)u_{\infty}(\tau)d\tau
+\int_{0}^{t}\Phi_{2}(t-\tau)\widehat{d}(\tau)d\tau+\Phi_{2}(t)d_{0}
\]

\noindent if and only if $\Gamma(s)G_{1}(s)$ is proper such that
\begin{equation}\label{b13}
\rho\left(I-\lim_{s\to\infty}\Gamma(s)G_{1}(s)\right)<1
\end{equation}

\noindent where particularly, $\Gamma(s)G_{1}(s)$ is nonsingular such that $U_{\infty}(s)=\mathcal{L}[u_{\infty}(t)]$ fulfills
\begin{equation}\label{b14}
U_{\infty}(s)=\left[\Gamma(s)G_{1}(s)\right]^{-1}\Gamma(s)\left[Y_{d}(s)-G_{2}(s)D(s)\right].
\end{equation}

\noindent Furthermore, the tracking objective (\ref{b1}) can be achieved if and only if $y_{d}(t)\in C^{1}_{q}[0,T]$ is trackable, where $U_{\infty}(s)=U_{d}(s)$ holds for $U_{d}(s)$ given by (\ref{b9}); and otherwise, $\lim_{k\to\infty}e_k(t)=e_{\infty}(t)\neq0$ holds, where $E_{\infty}(s)=\mathcal{L}[e_{\infty}(t)]$ satisfies
\begin{equation}\label{b15}
E_{\infty}(s)
=\left\{I-G_{1}(s)\left[\Gamma(s)G_{1}(s)\right]^{-1}\Gamma(s)\right\}\left[Y_{d}(s)-G_{2}(s)D(s)\right].
\end{equation}
\end{thm}

\begin{rem}\label{rem06}
From Theorem \ref{thm03}, we can find that the trackability of a specified trajectory is a necessary and sufficient condition for achieving the associated tracking objective in ILC of under-actuated systems. It particularly reveals that a continuous input sequence is generated by the ILC updating law (\ref{a12}) in the case of any trackable trajectory. This coincides with the trackability criterion established in Theorem \ref{thm01}. 
%
\end{rem}

For the system (\ref{b2}) in the over-actuated case (i.e., $q\leq p$), the results established in Theorem \ref{thm03} may no longer be applicable. Because of $q\leq p$, the convergence condition (\ref{b13}) for ILC may not hold although $\Gamma(s)G_{1}(s)$ is proper, where a straightforward consequence of the matrix theory \cite{hj:85} leads to
\begin{equation*}\label{}
\rho\left(I-\lim_{s\to\infty}\Gamma(s)G_{1}(s)\right)\geq1,\quad\forall q<p.
\end{equation*}

\noindent It clearly contradicts with (\ref{b13}). Despite this issue, it is possible for us to design $\Gamma(s)$ such that
\begin{equation}\label{b20}
\rho\left(I-\lim_{s\to\infty}G_{1}(s)\Gamma(s)\right)<1.
\end{equation}

\noindent By noting this condition for (\ref{b19}) and (\ref{a17}) and with Theorem \ref{thm02}, we employ an FCS-induced approach to develop the following theorem for the system (\ref{b2}) in the case $q\leq p$, which establishes a quite different ILC convergence result from Theorem \ref{thm03}.

\begin{thm}\label{thm04}
For the system (\ref{b2}) with $q\leq p$, let the conditions C1)--C4) be satisfied, and the ILC updating law (\ref{a12}) be applied with any initial input $U_{0}(s)=\mathcal{L}\left[u_{0}(t)\right]$ for $u_{0}(t)\in C_{p}[0,T]$ and any specified output trajectory $y_{d}(t)\in C^{1}_{q}[0,T]$ that satisfies the initial condition (\ref{b7}). Then $u_{k}(t)\in C_p[0,T]$, $\forall k\in\mathbb{Z}_{+}$ is such that $\lim_{k\to\infty}u_{k}(t)=u_{\infty}(t)$ holds for some $u_{\infty}(t)\in C_{p}[0,T]$, together with the tracking objective (\ref{b1}) being accomplished, if and only if $G_{1}(s)\Gamma(s)$ is proper and fulfills (\ref{b20}). Furthermore, $G_{1}(s)\Gamma(s)$ is nonsingular such that $U_{\infty}(s)=\mathcal{L}[u_{\infty}(t)]$ is dependent on the initial input $U_{0}(s)$ and forms a set given by
\begin{equation}\label{b21}
\aligned
\mathcal{U}_{\rm ILC}
&=\Big\{U_{\infty}(s)=\Gamma(s)\left[G_{1}(s)\Gamma(s)\right]^{-1}\left[Y_{d}(s)-G_{2}(s)D(s)\right]\\
&~~~~~~~~~~~~~~~+\widetilde{\Gamma}(s)U_{0}(s)\big|u_{0}(t)\in\mathbb{R}^{p}\Big\}
\endaligned
\end{equation}

\noindent where $\Gamma(s)=\left[\Gamma^{\tp}_{1}(s)~~\Gamma^{\tp}_{2}(s)\right]^{\tp}$ is denoted for $\Gamma_{1}(s)\in\R\F^{q\times q}(s)$ and $\Gamma_{2}(s)\in\R\F^{(p-q)\times q}(s)$ such that $\widetilde{\Gamma}(s)$ is given by
\[\aligned
&~~~~~~~~~~~~~~~~\widetilde{\Gamma}(s)
=\begin{bmatrix}
  \widetilde{\Gamma}_{11}(s) & \widetilde{\Gamma}_{12}(s)\\
  \widetilde{\Gamma}_{21}(s) & \widetilde{\Gamma}_{22}(s)
\end{bmatrix}~~\hbox{with}\\
&\left\{\aligned
\widetilde{\Gamma}_{11}(s)
&=G^{-1}_{11}(s)G_{12}(s)\Gamma_{2}(s)\left[G_{1}(s)\Gamma(s)\right]^{-1}G_{11}(s)\\
\widetilde{\Gamma}_{12}(s)
&=-G^{-1}_{11}(s)G_{12}(s)\left\{I-\Gamma_{2}(s)\left[G_{1}(s)\Gamma(s)\right]^{-1}G_{12}(s)\right\}\\
\widetilde{\Gamma}_{21}(s)
&=-\Gamma_{2}(s)\left[G_{1}(s)\Gamma(s)\right]^{-1}G_{11}(s)\\
\widetilde{\Gamma}_{22}(s)
&=I-\Gamma_{2}(s)\left[G_{1}(s)\Gamma(s)\right]^{-1}G_{12}(s).
\endaligned\right.
\endaligned
\]

\noindent In particular, $\mathcal{U}_{\rm ILC}=\mathcal{U}_{d}$ holds.
\end{thm}

\begin{rem}\label{rem07}
With Theorem \ref{thm04}, we reveal that for any specified trajectory, the ILC updating law (\ref{a12}) can be designed to realize the perfect tracking objective in the presence of over-actuated systems. It particularly indicates that by the selection of initial inputs, all inputs capable of generating the specified trajectory can be determined. This ILC tracking result is consistent with the trackability criterion developed in Theorem \ref{thm02}. In addition, the input induced from the ILC updating law (\ref{a12}) is continuous for every iteration if and only if the initial input is continuous since the specified output trajectory is trackable in ILC under the initial condition (\ref{b7}). 
\end{rem}

\begin{rem}\label{rem08}
In Theorems \ref{thm03} and \ref{thm04}, a unified condition is given to realize the convergence of ILC from the perspectives of both input and output, regardless of under-actuated or over-actuated systems. It ensures that for the ILC updating law (\ref{a12}) obtained with a feedback-based design method, the learned input $U_{\infty}(s)$ always exists and, particularly, is the same as the desired input for generating the trackable output trajectory in ILC. However, it is worth emphasizing that we make no assumption about the desired input in executing the ILC convergence analysis thanks to our introduced FCS-induced approach of ILC. This is quite different from classic convergence analysis approaches of ILC (see, e.g., \cite{akm:84,s:95,ack:93,sy:13}). Furthermore, our FCS-induced approach actually establishes the class of uniform convergence results for ILC by benefiting from Lemma \ref{lem08}.
\end{rem}

\begin{rem}\label{rem09}
If the initial condition (\ref{b7}) does not hold, and thus $y_{d}(t)$ is not trackable in ILC based on Theorems \ref{thm01} and \ref{thm02}, then by following the same way as the development of Theorems \ref{thm03} and \ref{thm04}, we can still establish the ILC convergence of both input and output, in spite of $q\geq p$ or $q\leq p$. It is worth emphasizing, however, that by (\ref{b17}), $u_{k}(t)\in\mathbb{R}^{p}$ may not be ensured. Further, the tracking objective (\ref{b1}) can still be realized in the case $q\leq p$, whereas it can not be accomplished in the case $q\geq p$. Namely, the use of the impulsive mechanism may be no longer effective in helping to achieve the perfect tracking objective for ILC in the presence of the initial shifts, which is different from \cite{pm:91}.
%
\end{rem}
%
%

\subsection{Further Discussions}

By \cite{bta:06,acm:07}, one of the practically important problems for the ILC systems is the robustness with respect to iteration-varying uncertainties. We thus proceed to develop the robustness of our trackability-based ILC results by reconsidering the system (\ref{b2}) in an uncertain form of
\begin{equation}\label{b33}
\aligned
Y_{k}(s)&=G_{1}(s)U_{k}(s)+G_{2}(s)D_{k}(s)\\
\endaligned
\end{equation}

\noindent where, in comparison with (\ref{b2}), $D_{k}(s)=D(s)+\Theta_{k}(s)$ holds and $\Theta_{k}(s)$ represents the iteration-varying uncertainty satisfying:
\begin{enumerate}
\item[C5)]
$\Theta_{k}(s)=\theta_{k}+\widehat{\Theta}_{k}(s)$ holds for some $\theta_{k}\in\mathbb{R}^{m}$ and $\widehat{\Theta}_{k}(s)\in\mathbb{RF}^{m\times1}$ such that $\left\|\theta_{k}\right\|\leq\beta_{\theta}$, $\forall k\in\mathbb{Z}_{+}$ and $\left\|\widehat{\theta}_{k}(t)\right\|\leq\beta_{\widehat{\theta}}$, $\forall t\in[0,T]$, $\forall k\in\mathbb{Z}_{+}$, where $\widehat{\theta}_{k}(t)\triangleq\mathcal{L}^{-1}\left[\widehat{\Theta}_{k}(s)\right]$ and $\beta_{\theta}$ and $\beta_{\widehat{\theta}}$ are some finite bounds.
\end{enumerate}

\noindent Then for Theorems \ref{thm03} and \ref{thm04}, we can show that they have certain robustness against iteration-varying uncertainties, as below.

\begin{cor}\label{cor03}
Consider the system (\ref{b33}) under the conditions C1), C2), C3) and C5). If the ILC updating law (\ref{a12}) is applied with any initial input $U_{0}(s)=\mathcal{L}\left[u_{0}(t)\right]$ for $u_{0}(t)\in C_{p}[0,T]$ and any specified output trajectory $y_{d}(t)\in C^{1}_{q}[0,T]$ that is trackable in ILC for the system (\ref{b2}), then the robust ILC tracking results can be established as follows.
\begin{enumerate}
\item
For $q\geq p$, let $\Gamma(s)G_{1}(s)$ be proper such that (\ref{b13}) holds. Then for $u_{\infty}(t)$ determined by (\ref{b14}),
\begin{equation}\label{b34}
\aligned
\limsup_{k \to \infty}\sup_{0\leq t\leq T}\left\|e_k(t)\right\|
&\leq \beta_e\\
\limsup_{k \to \infty}\sup_{0<t\leq T}\left\|u_k(t)-u_{\infty}(t)\right\|
&\leq \beta_u
\endaligned
\end{equation}

\noindent can be accomplished, where $\beta_e\geq0$ and $\beta_u\geq0$ are small bounds depending continuously on $\beta_{\theta}$ and $\beta_{\widehat{\theta}}$. In particular, when iteration-varying uncertainties disappear, i.e., $\beta_{\theta}\to0$ and $\beta_{\widehat{\theta}}\to0$, the same ILC convergence results as Theorem \ref{thm03} hold.

\item
For $q\leq p$, let the condition C4) hold, and $G_{1}(s)\Gamma(s)$ be proper such that (\ref{b20}) holds. Then the robust ILC tracking objective (\ref{b34}) can be achieved for some $u_{\infty}(t)$ defined by (\ref{b21}), and when iteration-varying uncertainties disappear, that is, $\beta_{\theta}\to0$ and $\beta_{\widehat{\theta}}\to0$, the same ILC convergence results as Theorem \ref{thm04} can be developed.
\end{enumerate}
\end{cor}

By Corollary \ref{cor03}, it indicates that like classic continuous-time ILC in, e.g., \cite{zm:21tcyb,sy:13}, the trackability-based ILC convergence results can be further extended to work robustly and effectively in the presence of iteration-varying uncertainties. This class of robust ILC convergence results may also be generalized to deal with iteration-varying uncertainties arising from plant models. Of special note is that the trackability-based ILC analysis gives a basic guarantee for the implementation of the ILC design and the robust convergence analysis.

Since most ILC results employ the time-domain descriptions \cite{acm:07}, next we revisit the time-domain system (\ref{a3}) (i.e., the state-space realization of the system (\ref{b2})), for which we particularly consider a commonly employed D-type ILC updating law as
\begin{equation}\label{eq32}
u_{k+1}(t)=u_k(t)+\Upsilon\dot{e}_k(t),\quad\forall t\in[0,T],\forall k\in\Z_+
\end{equation}

\noindent where $\Upsilon\in\mathbb{R}^{p\times q}$ is a constant gain matrix. Then with Theorems \ref{thm03} and \ref{thm04}, we can induce the following ILC convergence results.

\begin{cor}\label{cor04}
For the system (\ref{a3}), let the ILC updating law (\ref{eq32}) be applied under any initial input $u_{0}(t)\in C_{p}[0,T]$ and any specified trajectory $y_{d}(t)\in C^{1}_{q}[0,T]$. When $q\geq p$ (respectively, $q\leq p$), if $\rho(I-\Upsilon CB)<1$ (respectively, $\rho(I-CB\Upsilon)<1$), then $\lim_{k\to\infty}u_{k}(t)=u_{\infty}(t)$ and $\lim_{k\to\infty}y_{k}(t)=y_{\infty}(t)$ can be achieved for some $u_{\infty}(t)\in\mathbb{R}^{p}$ and $y_{\infty}(t)\in\mathbb{R}^{q}$. Furthermore, the tracking objective (\ref{b1}) can be accomplished, together with giving $u_{k}(t)\in C_p[0,T]$, $\forall k\in\mathbb{Z}_{+}$, if and only if $y_{d}(t)\in C^{1}_{q}[0,T]$ is trackable.
\end{cor}

With Corollary \ref{cor04}, we can see that the trackability-based ILC convergence results are particularly applicable for the classical D-type ILC. But, differently, Corollary \ref{cor04} reveals that a unified condition can be obtained to ensure the ILC convergence from the perspectives of both input and output, regardless of under-actuated or over-actuated systems. This can not be gained with typical ILC analysis methods (see, e.g., \cite{akm:84,zm:21tcyb,pm:91,so:91,sy:13,lch:05}).

\subsection{FCS-Induced Convergence Analysis}

Next, we give the proofs of Lemma \ref{lem06} and Theorems \ref{thm03} and \ref{thm04} by applying Lemmas \ref{lem08} and \ref{lem07} and using the frequency-domain analysis method.

\begin{IEEEproof}[Proof of Lemma \ref{lem06}]
For $\Gamma(s)\in\R\F^{p\times q}(s)$, let $\gamma(s)$ be the monic least common denominator of all its nonzero entries. We without any loss of generality represent $\gamma(s)$ as $\gamma(s)=\sum_{i=0}^{m}\gamma_{i}s^{i}$ for some $m\in\mathbb{Z}_{+}$, some $\gamma_{i}\in\mathbb{R}$, $\forall i=0$, $1$, $\cdots$, $m-1$, and $\gamma_{m}=1$. Then we can write $\Gamma(s)$ in the form of
\begin{equation}\label{a1}
\Gamma(s)=\gamma^{-1}(s)\Xi(s)
\end{equation}

\noindent where $\Xi(s)\in\R\P^{p\times q}(s)$ is a polynomial matrix. Then let $\Xi(s)$ be of the form (see, e.g., \cite[(2.6), p. 528]{am:06})
\begin{equation}\label{a2}
\aligned
\Xi(s)
&=\Gamma_{0}s^{l}+\Gamma_{1}s^{l-1}+\cdots+\Gamma_{l-1}s+\Gamma_{l}\\
&\triangleq\Gamma_{0}s^{l}+\text{lower degree terms}.
\endaligned
\end{equation}

\noindent where $l\in\mathbb{Z}_{+}$ and $\Gamma_{i}\in\mathbb{R}^{p\times q}$, $\forall i=0$, $1$, $\cdots$, $l$ with $\Gamma_{0}\neq0$. The preliminary results of (\ref{a1}) and (\ref{a2}) help us to deduce that any of the conditions 1)--3) holds if and only if $l=m+1$. Inspired by this fact, we next consider the conditions 1)--3) separately.

With (\ref{a27}), (\ref{a28}), (\ref{a1}), and (\ref{a2}), we can write $\Gamma(s)G_{1}(s)$ as
\begin{equation}\label{a8}
\aligned
\Gamma(s)G_1(s)
&=\left[\gamma(s)\alpha(s)\right]^{-1}\Xi(s)Q(s)\\
&=\left(s^{n+m}+\hbox{lower degree terms}\right)^{-1}\\
&~~~\times\left[\Gamma_{0}\Phi_{1}(0)s^{n+l-1}+\hbox{lower degree terms}\right].
\endaligned
\end{equation}

\noindent With (\ref{a8}), it follows straightforwardly that $\Gamma(s)G_{1}(s)$ is proper if and only if $l=m+1$, and consequently,
$\lim_{s\to\infty}\Gamma(s)G_1(s)=\Gamma_{0}\Phi_{1}(0)$. For the same reason as (\ref{a8}), we can also arrive at
\begin{equation}\label{a9}
\aligned
G_1(s)\Gamma(s)
&=\left[\alpha(s)\gamma(s)\right]^{-1}Q(s)\Xi(s)\\
&=\left(s^{n+m}+\hbox{lower degree terms}\right)^{-1}\\
&~~~\times\left[\Phi_{1}(0)\Gamma_{0}s^{n+l-1}+\hbox{lower degree terms}\right]
\endaligned
\end{equation}

\noindent from which $G_{1}(s)\Gamma(s)$ is proper if and only if $l=m+1$. Hence, the use of (\ref{a9}) gives $\lim_{s\to\infty}G_1(s)\Gamma(s)=\Phi_{1}(0)\Gamma_{0}$. In addition, we can leverage (\ref{a1}) and (\ref{a2}) to obtain
\begin{equation}\label{a10}
\aligned
s^{-1}\Gamma(s)
&=\left[s\gamma(s)\right]^{-1}\Xi(s)\\
&=\left(s^{m+1}+\hbox{lower degree terms}\right)^{-1}\\
&~~~\times\left(\Gamma_{0}s^{l}+\hbox{lower degree terms}\right)
\endaligned
\end{equation}

\noindent which obviously guarantees that $s^{-1}\Gamma(s)$ is proper if and only if $l=m+1$. Then as a consequence of (\ref{a10}), (\ref{b16}) is immediate.

To proceed, we note $y_{d}(t)\in C^{1}_{q}[0,T]$, and thus have $sE_{k}(s)=\mathcal{L}\left[\dot{e}_{k}(t)\right]+e_{k}(0)$. Then we incorporate the condition 3) to get
\begin{equation}\label{a11}
\aligned
U_{k+1}(s)&=U_{k}(s)+\left[s^{-1}\Gamma(s)\right]\left[sE_{k}(s)\right]\\
&=U_{k}(s)+\Gamma_{0}\mathcal{L}\left[\dot{e}_{k}(t)\right]
+\widehat{\Gamma}(s)\mathcal{L}\left[\dot{e}_{k}(t)\right]\\
&~~~+\widehat{\Gamma}(s)e_{k}(0)+\Gamma_{0}e_{k}(0).
\endaligned
\end{equation}

\noindent By taking the inverse Laplace transform on both sides of (\ref{a11}), we can directly derive (\ref{b17}). Then in view of (\ref{b5}), we can clearly conclude from (\ref{b17}) that $u_{k}(t)\in C_p[0,T]$, $\forall k\in\mathbb{Z}_{+}$ if and only if $\Gamma_{0}e_{k}(0)\delta(t)=0$, $\forall k\in\mathbb{Z}_{+}$, i.e., $\Gamma_{0}e_{k}(0)=0$, $\forall k\in\mathbb{Z}_{+}$. Hence, (\ref{b18}) is obtained.
\end{IEEEproof}

\begin{IEEEproof}[Proof of Theorem \ref{thm03}]
Since the initial condition (\ref{b7}) holds, we have $u_{k}(t)\in C_p[0,T]$, $\forall k\in\mathbb{Z}_{+}$ from Lemma \ref{lem06}. Then owing to $q\geq p$, we consider Lemma \ref{lem08} for (\ref{a17}) and can arrive at that the following three results are equivalent:
\begin{enumerate}
\item
$\lim_{k\to\infty}u_{k}(t)=u_{\infty}(t)\in C_p[0,T]$ holds with its limit being approached uniformly on $[0,T]$, which together with (\ref{b5}) thus results in $\lim_{k\to\infty}y_{k}(t)=y_{\infty}(t)\in C^{1}_q[0,T]$;

\item
$\left\{u_{k}(t):k\in\Z_{+}\right\}$ is an FCS;

\item
$\Gamma(s)G_{1}(s)$ is proper such that (\ref{b13}) holds.
\end{enumerate}

\noindent To proceed, we can easily leverage Lemmas \ref{lem01} and \ref{lem06} to validate that $\lim_{s\to\infty}\Gamma(s)G_{1}(s)=\Gamma_{0}\Phi_{1}(0)$ holds. Then as an immediate consequence of (\ref{b13}), $\Gamma_{0}\Phi_{1}(0)$ is nonsingular. In the same way as the proof of Lemma \ref{lem04}, we can further obtain that $\Gamma(s)G_{1}(s)$ is nonsingular. This, together with (\ref{a16}), leads to (\ref{b14}) directly.

Next, we prove the equivalence between the tracking objective (\ref{b1}) and the trackability of $y_{d}(t)$.

{\it Necessity:} If the tracking objective (\ref{b1}) is achieved, namely, $\lim_{k\to\infty}Y_k(s)=Y_{d}(s)$, then by $\lim_{k\to\infty}U_k(s)=U_{\infty}(s)$, it follows immediately from (\ref{b2}) that
\begin{equation*}\label{}
\aligned
G_{1}(s)U_{\infty}(s)
&=Y_{\infty}(s)-G_{2}(s)D(s)
=Y_d(s)-G_{2}(s)D(s)
\endaligned
\end{equation*}

\noindent where $Y_{\infty}(s)\triangleq\lim_{k\to\infty}Y_{k}(s)$. Namely, $U_{\infty}(s)$ is a solution of the algebraic equation (\ref{b3}). Then by Definition \ref{defi2}, $y_d(t)$ is trackable.

{\it Sufficiency:} If $y_d(t)$ is trackable, then according to Theorem \ref{thm01}, the algebraic equation (\ref{b3}) has a unique solution $U_d(s)$ shown by (\ref{b9}). This, together with (\ref{a16}), yields
\begin{equation}\label{a87}
U_{k+1}(s)=[I-\Gamma(s)G_1(s)]U_{k}(s)+\Gamma(s)G_1(s)U_d(s)
\end{equation}

\noindent by which the use of $\lim_{k \to \infty}U_k(s)=U_{\infty}(s)$ leads to
\begin{equation}
\label{a88}
\Gamma(s)G_1(s)U_{\infty}(s)=\Gamma(s)G_1(s)U_d(s).
\end{equation}

\noindent Because $\Gamma(s)G_1(s)$ is nonsingular, we can apply (\ref{a88}) to arrive at $U_{\infty}(s)=U_d(s)$. As a consequence, we also have
\begin{equation*}\label{}
\aligned
Y_{\infty}(s)
&=G_{1}(s)U_{\infty}(s)+G_{2}(s)D(s)\\
&=G_{1}(s)U_{d}(s)+G_{2}(s)D(s)\\
&=Y_{d}(s)
\endaligned
\end{equation*}

\noindent namely, the tracking objective (\ref{b1}) can be achieved. 
%

Besides, the abovementioned necessary and sufficient results guarantee that when $y_d(t)$ is not trackable, or equivalently, the tracking objective (\ref{b1}) does not hold, $\lim_{k\to\infty}e_k(t)=e_{\infty}(t)\neq0$ is thus obvious, where the use of (\ref{b14}) results in
\[
\aligned
E_{\infty}(s)
&=\lim_{k \to \infty}E_k(s)\\
&=Y_d(s)-G_1(s)U_{\infty}(s)-G_2(s)D(s)\\
&=\left\{I-G_1(s)[\Gamma(s)G_1(s)]^{-1}\Gamma(s)\right\}\left[Y_d(s)-G_2(s)D(s)\right]\\
&\neq0
\endaligned
\]

\noindent i.e., (\ref{b15}) holds.
\end{IEEEproof}

\begin{IEEEproof}[Proof of Theorem \ref{thm04}]
Thanks to the initial condition (\ref{b7}), it follows that $u_{k}(t)\in C_p[0,T]$, $\forall k\in\mathbb{Z}_{+}$ holds based on Lemma \ref{lem06}, and that $y_{d}(t)\in C^{1}_q[0,T]$ is trackable in ILC by Theorem \ref{thm02}. Next, we show the necessity and sufficiency separately.

{\it Necessity:} Because the tracking objective (\ref{b1}) is realized, it is direct that $\lim_{k\to\infty}e_{k}(t)=0$, $\forall t\in[0,T]$. By considering Lemma \ref{lem08} for (\ref{b19}) and (\ref{a17}) and applying Lemma \ref{lem01} for $u_{k}(t)\in C_p[0,T]$, $\forall k\in\mathbb{Z}_{+}$, we can develop that if $\lim_{k\to\infty}u_{k}(t)=u_{\infty}(t)$, together with the tracking objective (\ref{b1}) being achieved, then $G_{1}(s)\Gamma(s)$ is proper such that (\ref{b20}) holds.

{\it Sufficiency:} If $G_{1}(s)\Gamma(s)$ is proper such that (\ref{b20}) holds, then by following the same lines as adopted in the proof of Theorem \ref{thm03}, we can deduce that $G_{1}(s)\Gamma(s)$ is nonsingular, where we have $\lim_{s\to\infty}G_{1}(s)\Gamma(s)=\Phi_{1}(0)\Gamma_{0}$ and $\Phi_{1}(0)\Gamma_{0}$ is also nonsingular. With this fact and by Lemma \ref{lem05}, we denote a structured matrix $\Omega(s)\in\R\F^{p\times p}(s)$ in the form of
\begin{equation*}\label{}
\Omega(s)=\begin{bmatrix}\Omega_{11}(s)&\Omega_{12}(s)\\
\Omega_{21}(s)&\Omega_{22}(s)\end{bmatrix}
\end{equation*}

\noindent where four block matrices involved in $\Omega(s)$ are given by
\begin{equation*}\label{}
\aligned
\Omega_{11}(s)&=s\left[G_1(s)\Gamma(s)\right]^{-1}G_{11}(s)\\
\Omega_{12}(s)&=s\left[G_1(s)\Gamma(s)\right]^{-1}G_{12}(s)\\
\Omega_{21}(s)&=-\Gamma_{2}(s)\left[G_1(s)\Gamma(s)\right]^{-1}G_{11}(s)\\
\Omega_{22}(s)&=I-\Gamma_{2}(s)\left[G_1(s)\Gamma(s)\right]^{-1}G_{12}(s).
\endaligned
\end{equation*}

\noindent We can validate that $\Omega(s)$ is nonsingular, and proper due to
\begin{equation*}\label{}
\aligned
&\lim_{s\to\infty}\Omega(s)\\
&=\begin{bmatrix}
\left[\Phi_{1}(0)\Gamma_{0}\right]^{-1}\Phi_{1,1}(0) & \left[\Phi_{1}(0)\Gamma_{0}\right]^{-1}\Phi_{1,2}(0) \\
-\Gamma_{0,2}\left[\Phi_{1}(0)\Gamma_{0}\right]^{-1}\Phi_{1,1}(0) & I-\Gamma_{0,2}\left[\Phi_{1}(0)\Gamma_{0}\right]^{-1}\Phi_{1,2}(0)
\end{bmatrix}
\endaligned
\end{equation*}

\noindent where $\Gamma_{0,2}\in\mathbb{R}^{(p-q)\times q}$, together with $\Gamma_{0,1}\in\mathbb{R}^{q\times q}$, is such that $\Gamma_{0}=\left[\Gamma_{0,1}^{\tp}~~\Gamma_{0,2}^{\tp}\right]^{\tp}$. Simultaneously, the inverse matrix $\Psi(s)\triangleq\Omega^{-1}(s)\in\R\F^{p\times p}(s)$ satisfies
\begin{equation*}\label{}
\Psi(s)
=\begin{bmatrix}\Psi_{11}(s)&\Psi_{12}(s)\\
\Psi_{21}(s)&\Psi_{22}(s)\end{bmatrix}
\end{equation*}

\noindent where four block matrices taking the form of
\begin{equation*}\label{}
\aligned
\Psi_{11}(s)&=s^{-1}\Gamma_1(s),&
\Psi_{12}(s)&=-G^{-1}_{11}(s)G_{12}(s)\\
\Psi_{21}(s)&=s^{-1}\Gamma_{2}(s),&
\Psi_{22}(s)&=I
\endaligned
\end{equation*}

\noindent are such that $\Psi(s)$ is also proper thanks to
\begin{equation*}\label{}
\lim_{s\to\infty}\Psi(s)
=\begin{bmatrix}
\Gamma_{0,1} & -\Phi_{1,1}^{-1}(0)\Phi_{1,2}(0) \\
\Gamma_{0,2} & I
\end{bmatrix}.
\end{equation*}

To proceed, we employ $\Omega(s)$ to propose a nonsingular linear transformation as
\begin{equation*}\label{}
\Omega(s)U_{k}(s)=U_{k}^{\star}(s)\triangleq\begin{bmatrix}U_{k,1}^{\star}(s)\\U_{k,2}^{\star}(s)\end{bmatrix}
\end{equation*}

\noindent where we correspondingly denote $u^{\star}_k(t)=\L^{-1}\left[U^{\star}_k(s)\right]\in\mathbb{R}^{p}$, $u^{\star}_{k,1}(t)=\L^{-1}\left[U^{\star}_{k,1}(s)\right]\in\mathbb{R}^{q}$, and $u^{\star}_{k,2}(t)=\L^{-1}\left[U^{\star}_{k,2}(s)\right]\in\mathbb{R}^{p-q}$. Due to that $u_{k}(t)\in C_p[0,T]$, $\forall k\in\mathbb{Z}_{+}$ and $\Omega(s)$ is proper, it follows that for all $k\in\mathbb{Z}_{+}$, we have $u^{\star}_{k}(t)\in C_{p}[0,T]$, $u^{\star}_{k,1}(t)\in C_{q}[0,T]$, and $u^{\star}_{k,2}(t)\in C_{p-q}[0,T]$. Since we can easily verify
\begin{equation}\label{a15}
\Omega(s)\Gamma(s)=\begin{bmatrix}sI\\0\end{bmatrix},\quad
G_{1}(s)\Psi(s)=\begin{bmatrix}s^{-1}G_{1}(s)\Gamma(s)&0\end{bmatrix}
\end{equation}

\noindent we again consider (\ref{a16}) and can arrive at
\begin{equation*}\label{}
\aligned
U_{k+1}^{\star}(s)
&=\Omega(s)U_{k+1}(s)\\
&=\left\{I-\left[\Omega(s)\Gamma(s)\right]\left[G_{1}(s)\Psi(s)\right]\right\}\Omega(s)U_{k}(s)\\
&~~~+\left[\Omega(s)\Gamma(s)\right]\left[Y_d(s)-G_{2}(s)D(s)\right]\\
&=\left\{I-\begin{bmatrix}G_{1}(s)\Gamma(s)&0\\0&0\end{bmatrix}\right\}U_{k}^{\star}(s)\\
&~~~+\begin{bmatrix}sI\\0\end{bmatrix}\left[Y_d(s)-G_{2}(s)D(s)\right]
\endaligned
\end{equation*}

\noindent and, consequently, $U_{k,1}^{\star}(s)$ and $U_{k,2}^{\star}(s)$ are decoupled from each other such that
\begin{equation}\label{b22}
\aligned
\begin{bmatrix}U_{k+1,1}^{\star}(s)\\U_{k+1,2}^{\star}(s)\end{bmatrix}
&=\begin{bmatrix}I-G_{1}(s)\Gamma(s)&0\\0&I\end{bmatrix}\begin{bmatrix}U_{k,1}^{\star}(s)\\U_{k,2}^{\star}(s)\end{bmatrix}\\
&~~~+\begin{bmatrix}s\left[Y_d(s)-G_{2}(s)D(s)\right]\\0\end{bmatrix},\quad\forall k\in\mathbb{Z}_{+}.
\endaligned
\end{equation}

\noindent A direct consequence of (\ref{b22}) is that $U_{k+1,2}^{\star}(s)=U_{k,2}^{\star}(s)$, $\forall k\in\mathbb{Z}_{+}$, namely, $U_{k,2}^{\star}(s)$ is iteration-independent such that
\begin{equation}\label{b23}
U_{k,2}^{\star}(s)
\equiv U_{0,2}^{\star}(s)
=\begin{bmatrix}
\Omega_{21}(s) & \Omega_{22}(s)
\end{bmatrix}U_0(s),\quad\forall k\in\mathbb{Z}_{+}.
\end{equation}

\noindent In addition, the use of (\ref{b22}) leads to
\begin{equation}\label{b24}
\aligned
U^{\star}_{k+1,1}(s)
&=\left[I-G_{1}(s)\Gamma(s)\right]U^{\star}_{k,1}(s)\\
&~~~+s\left[Y_d(s)-G_{2}(s)D(s)\right],\quad\forall k\in\mathbb{Z}_{+}
\endaligned
\end{equation}

\noindent which immediately yields
\begin{equation}\label{b25}
\aligned
U^{\star}_{k+2,1}(s)-U^{\star}_{k+1,1}(s)
&=\left[I-G_{1}(s)\Gamma(s)\right]\\
&~~~\times\left[U^{\star}_{k+1,1}(s)-U^{\star}_{k,1}(s)\right],\quad\forall k\in\mathbb{Z}_{+}.
\endaligned
\end{equation}

\noindent Since $G_{1}(s)\Gamma(s)$ is proper and (\ref{b20}) holds, $\left\{u^{\star}_{k,1}(t):k\in\Z_+\right\}$ is an FCS by applying Lemma \ref{lem08} to (\ref{b25}). Thus, there exists some function $u^{\star}_{\infty,1}(t)\in C_{q}[0,T]$ such that $\lim_{k\to\infty}u^{\star}_{k,1}(t)=u^{\star}_{\infty,1}(t)$ in view of $u^{\star}_{k,1}(t)\in C_{q}[0,T]$, $\forall k\in\mathbb{Z}_{+}$. This, together with (\ref{b24}) and the nonsingularity of $G_{1}(s)\Gamma(s)$, implies 
\begin{equation}\label{b26}
\aligned
U^{\star}_{\infty,1}(s)
&\triangleq\lim_{k\to\infty}U^{\star}_{k,1}(s)\\
&=\left[G_{1}(s)\Gamma(s)\right]^{-1}s\left[Y_d(s)-G_{2}(s)D(s)\right].
\endaligned
\end{equation}

\noindent From (\ref{b23}) and (\ref{b26}), it is immediate to derive
\begin{equation}\label{b27}
\aligned
U^{\star}_{\infty}(s)
&\triangleq\lim_{k\to\infty}U^{\star}_{k}(s)\\
&=\begin{bmatrix}\left[G_{1}(s)\Gamma(s)\right]^{-1}s\left[Y_d(s)-G_{2}(s)D(s)\right]\\
\begin{bmatrix}
\Omega_{21}(s) & \Omega_{22}(s)
\end{bmatrix}U_0(s)\end{bmatrix}.
\endaligned
\end{equation}

\noindent For (\ref{b21}), we can easily verify
\[
\widetilde{\Gamma}(s)=\begin{bmatrix}
\Psi_{12}(s)\\\Psi_{22}(s)
\end{bmatrix}\begin{bmatrix}
\Omega_{21}(s) & \Omega_{22}(s)
\end{bmatrix}
\]

\noindent and then the use of (\ref{b27}), together with $U^{\star}_k(s)=\Omega(s)U_k(s)$ and $\Omega^{-1}(s)=\Psi(s)$, leads to
\begin{equation}\label{b28}
\aligned
U_{\infty}(s)
&\triangleq\lim_{k\to\infty}U_{k}(s)\\
&=\Psi(s)\begin{bmatrix}\left[G_{1}(s)\Gamma(s)\right]^{-1}s\left[Y_d(s)-G_{2}(s)D(s)\right]\\
\begin{bmatrix}
\Omega_{21}(s) & \Omega_{22}(s)
\end{bmatrix}U_0(s)\end{bmatrix}\\
&=\begin{bmatrix}
\Psi_{11}(s)\\\Psi_{21}(s)
\end{bmatrix}\left[G_{1}(s)\Gamma(s)\right]^{-1}s\left[Y_d(s)-G_{2}(s)D(s)\right]\\
&~~~+\begin{bmatrix}
\Psi_{12}(s)\\\Psi_{22}(s)
\end{bmatrix}\begin{bmatrix}
\Omega_{21}(s) & \Omega_{22}(s)
\end{bmatrix}U_{0}(s)\\
&=\Gamma(s)\left[G_{1}(s)\Gamma(s)\right]^{-1}\left[Y_d(s)-G_{2}(s)D(s)\right]
+\widetilde{\Gamma}(s)U_{0}(s)
\endaligned
\end{equation}

\noindent namely, (\ref{b21}) holds. Furthermore, we incorporate (\ref{a15}) into (\ref{b28}), and can validate
\begin{equation*}\label{}
\aligned
G_1(s)U_{\infty}(s)
&=G_1(s)\Psi(s)\begin{bmatrix}\left[G_{1}(s)\Gamma(s)\right]^{-1}s\left[Y_d(s)-G_{2}(s)D(s)\right]\\
\begin{bmatrix}
\Omega_{21}(s) & \Omega_{22}(s)
\end{bmatrix}U_0(s)\end{bmatrix}\\
&=Y_d(s)-G_{2}(s)D(s)
\endaligned
\end{equation*}

\noindent which ensures
\[\aligned
\lim_{k\to\infty}E_{k}(s)
&=Y_{d}(s)-\lim_{k\to\infty}Y_{k}(s)\\
&=Y_{d}(s)-\left[G_{1}(s)U_{\infty}(s)+G_{2}(s)D(s)\right]\\
&=0
\endaligned
\]

\noindent and, consequently, the tracking objective (\ref{b1}) can be realized.

With the above necessary and sufficient result, we next show $\mathcal{U}_{\rm ILC}=\mathcal{U}_{d}$ by adopting two steps.

{\it i): $\mathcal{U}_{\rm ILC}\subseteq\mathcal{U}_{d}$.} For any $U_{\infty}(s)\in\mathcal{U}_{\rm ILC}$, we take
\[\aligned
U_{d,2}(s)
&=\begin{bmatrix}
\Omega_{21}(s) & \Omega_{22}(s)
\end{bmatrix}U_0(s)\\
&~~~+\Gamma_{2}(s)\left[G_{1}(s)\Gamma(s)\right]^{-1}\left[Y_d(s)-G_{2}(s)D(s)\right]
\endaligned\]

\noindent which, together with (\ref{b28}), results in
\begin{equation*}\label{}
\aligned
U_{\infty}(s)
&=\Psi(s)\\
&~~~\times\begin{bmatrix}\left[G_{1}(s)\Gamma(s)\right]^{-1}s\left[Y_d(s)-G_{2}(s)D(s)\right]\\
U_{d,2}(s)-\Gamma_{2}(s)\left[G_{1}(s)\Gamma(s)\right]^{-1}\left[Y_d(s)-G_{2}(s)D(s)\right]\end{bmatrix}\\
&=\begin{bmatrix}
G_{11}^{-1}(s)\left[Y_d(s)-G_{2}(s)D(s)-G_{12}(s)U_{d,2}(s)\right]\\
U_{d,2}(s)
\end{bmatrix}
\endaligned
\end{equation*}

\noindent namely, $U_{\infty}(s)\in\mathcal{U}_{d}$. As a consequence, we have $\mathcal{U}_{\rm ILC}\subseteq\mathcal{U}_{d}$.

{\it ii): $\mathcal{U}_{\rm ILC}\supseteq\mathcal{U}_{d}$.} For any $U_{d}(s)\in\mathcal{U}_{d}$, let us take
\[\aligned
U_{0}(s)
&=\begin{bmatrix}
-G_{11}^{-1}(s)G_{12}(s)\\
I
\end{bmatrix}\Big\{U_{d,2}(s)\\
&~~~-\Gamma_{2}(s)\left[G_{1}(s)\Gamma(s)\right]^{-1}\left[Y_d(s)-G_{2}(s)D(s)\right]\Big\}
\endaligned\]

\noindent and, consequently, we can use $\Omega_{21}(s)\Psi_{12}(s)+\Omega_{22}(s)\Psi_{22}(s)=I$ to derive
\[\aligned
\widetilde{\Gamma}(s)U_{0}(s)
&=\begin{bmatrix}\Psi_{12}(s)\\\Psi_{22}(s)\end{bmatrix}
\begin{bmatrix}\Omega_{21}(s)&\Omega_{22}(s)\end{bmatrix}\\
&~~~\times\begin{bmatrix}-G_{11}^{-1}(s)G_{12}(s)\\I\end{bmatrix}\Big\{U_{d,2}(s)\\
&~~~-\Gamma_{2}(s)\left[G_{1}(s)\Gamma(s)\right]^{-1}\left[Y_d(s)-G_{2}(s)D(s)\right]\Big\}\\
&=\begin{bmatrix}-G_{11}^{-1}(s)G_{12}(s)\\I\end{bmatrix}\Big\{U_{d,2}(s)\\
&~~~-\Gamma_{2}(s)\left[G_{1}(s)\Gamma(s)\right]^{-1}\left[Y_d(s)-G_{2}(s)D(s)\right]\Big\}
\endaligned\]

\noindent which yields
\[\aligned
\begin{bmatrix}-G^{-1}_{11}(s)G_{12}(s)\\I\end{bmatrix}U_{d,2}(s)
&=\begin{bmatrix}-G^{-1}_{11}(s)G_{12}(s)\\I\end{bmatrix}\Gamma_{2}(s)
\left[G_{1}(s)\Gamma(s)\right]^{-1}\left[Y_d(s)-G_{2}(s)D(s)\right]\\
&~~~+\widetilde{\Gamma}(s)U_{0}(s).
\endaligned\]

\noindent This, together with (\ref{b12}), leads to
\[\aligned
U_{d}(s)
&=\begin{bmatrix}G^{-1}_{11}(s)\left[Y_{d}(s)-G_{2}(s)D(s)\right]\\0\end{bmatrix}\\
&~~~+\begin{bmatrix}-G^{-1}_{11}(s)G_{12}(s)\\I\end{bmatrix}\Gamma_{2}(s)
\left[G_{1}(s)\Gamma(s)\right]^{-1}\\
&~~~\times\left[Y_d(s)-G_{2}(s)D(s)\right]+\widetilde{\Gamma}(s)U_{0}(s)\\
&=\Gamma(s)\left[G_{1}(s)\Gamma(s)\right]^{-1}\left[Y_d(s)-G_{2}(s)D(s)\right]
+\widetilde{\Gamma}(s)U_{0}(s)
\endaligned\]

\noindent which implies $U_{d}(s)\in\mathcal{U}_{\rm ILC}$. Thus, $\mathcal{U}_{\rm ILC}\supseteq\mathcal{U}_{d}$ is immediate.

With the steps i) and ii), we can arrive at $\mathcal{U}_{\rm ILC}=\mathcal{U}_{d}$.
\end{IEEEproof}

\section{Simulation Examples}\label{sec4}

\begin{figure*}[!t]
\centering
\includegraphics[width=0.32\hsize]{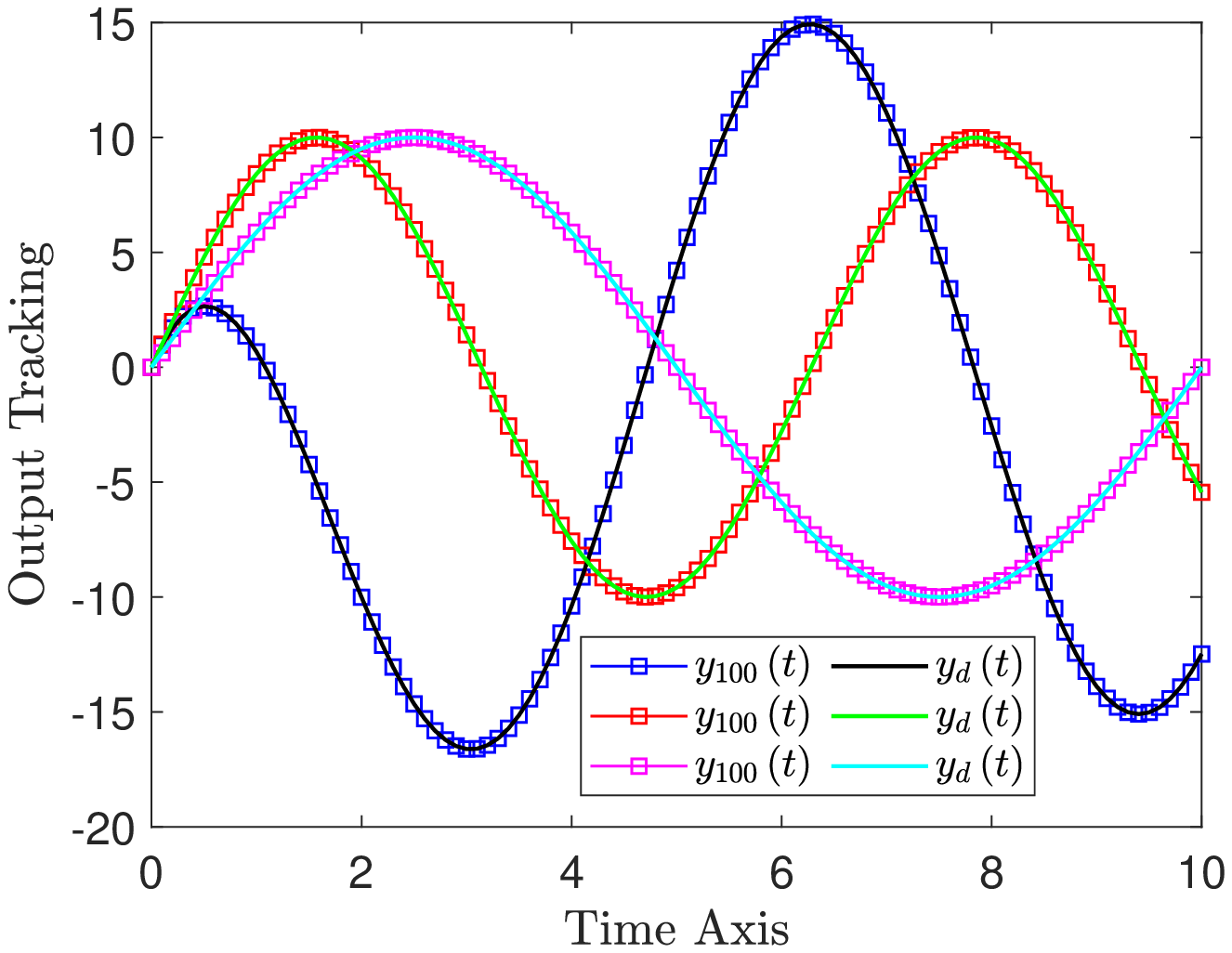}~\includegraphics[width=0.32\hsize]{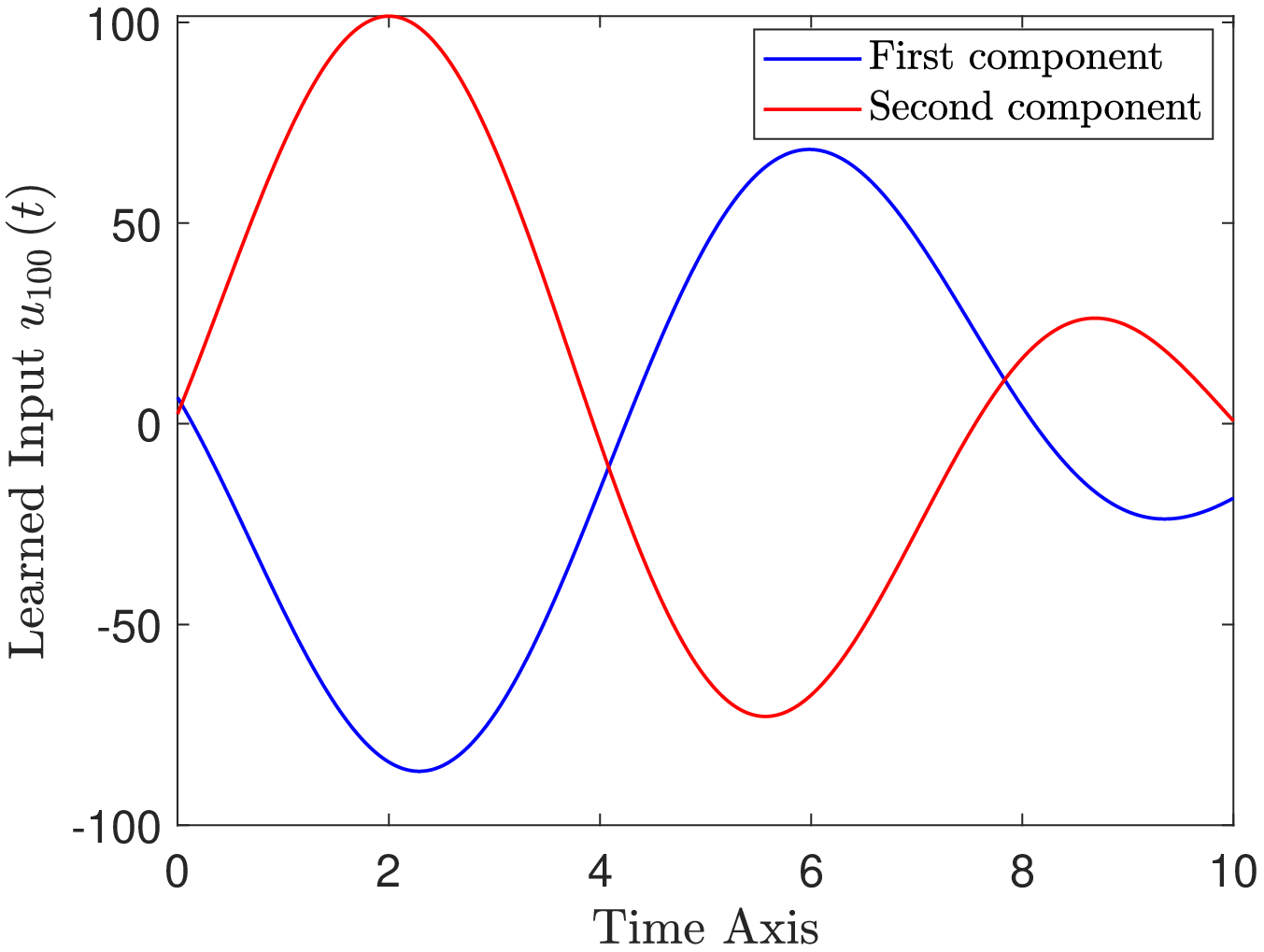}~\includegraphics[width=0.32\hsize]{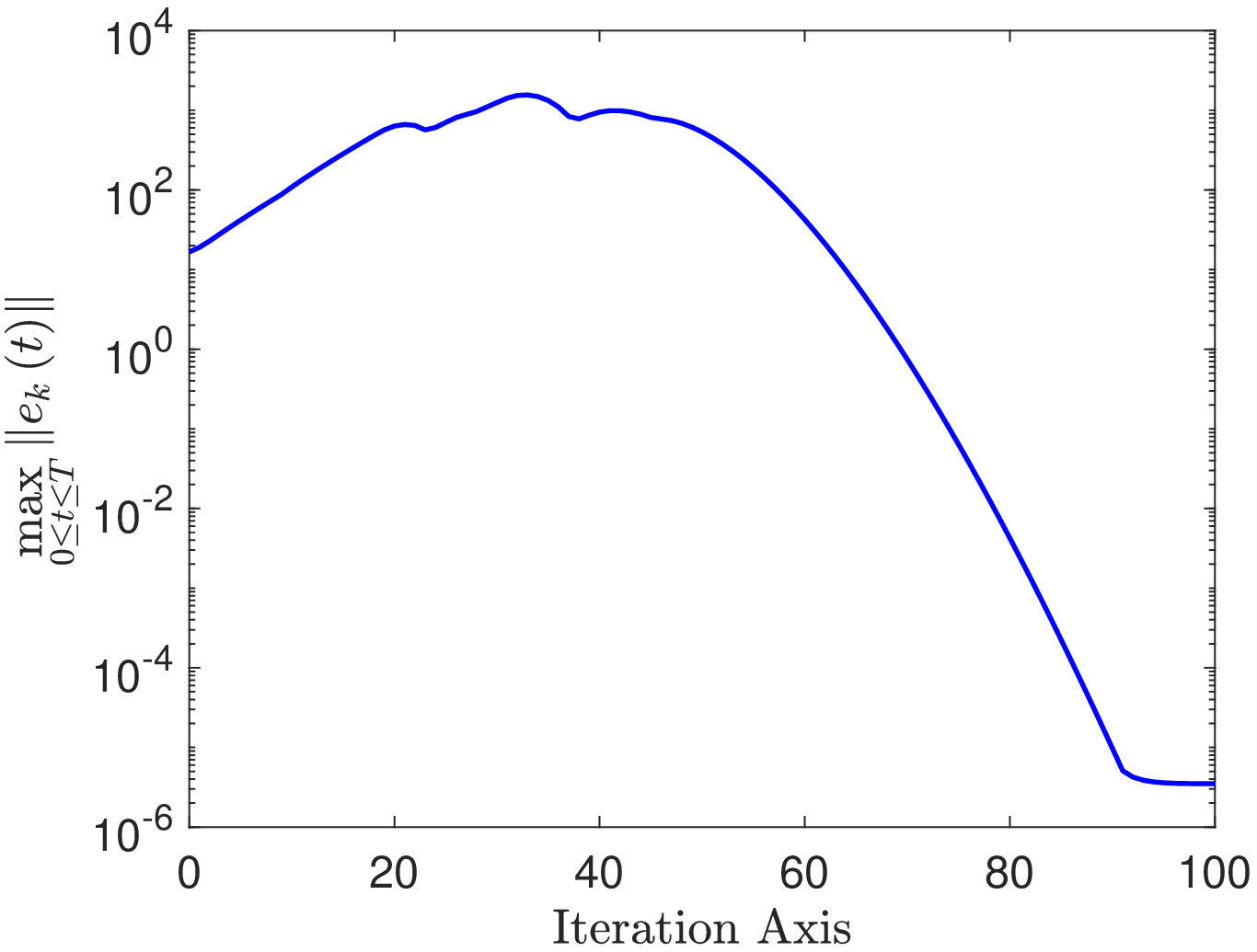}\\
\includegraphics[width=0.32\hsize]{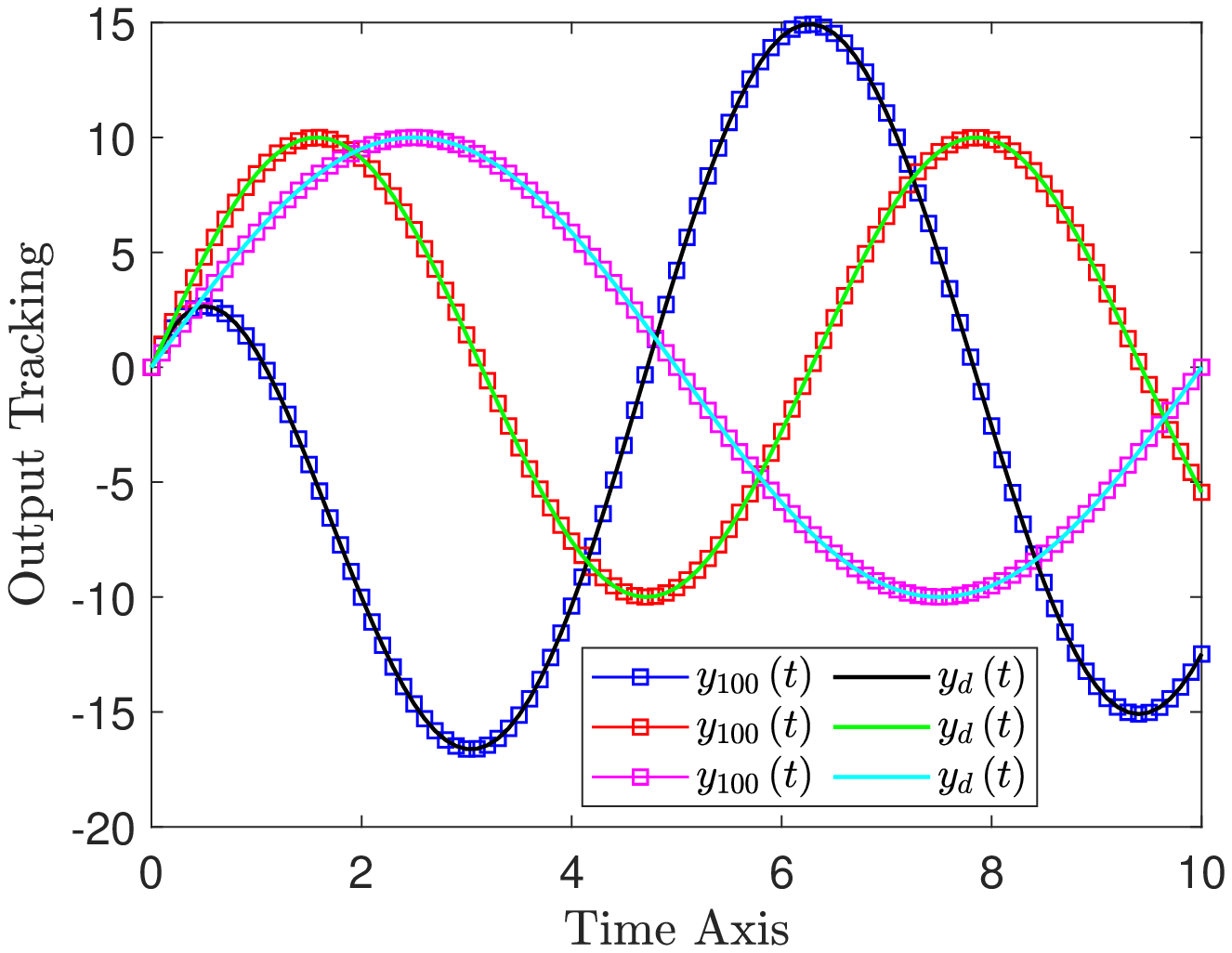}~\includegraphics[width=0.32\hsize]{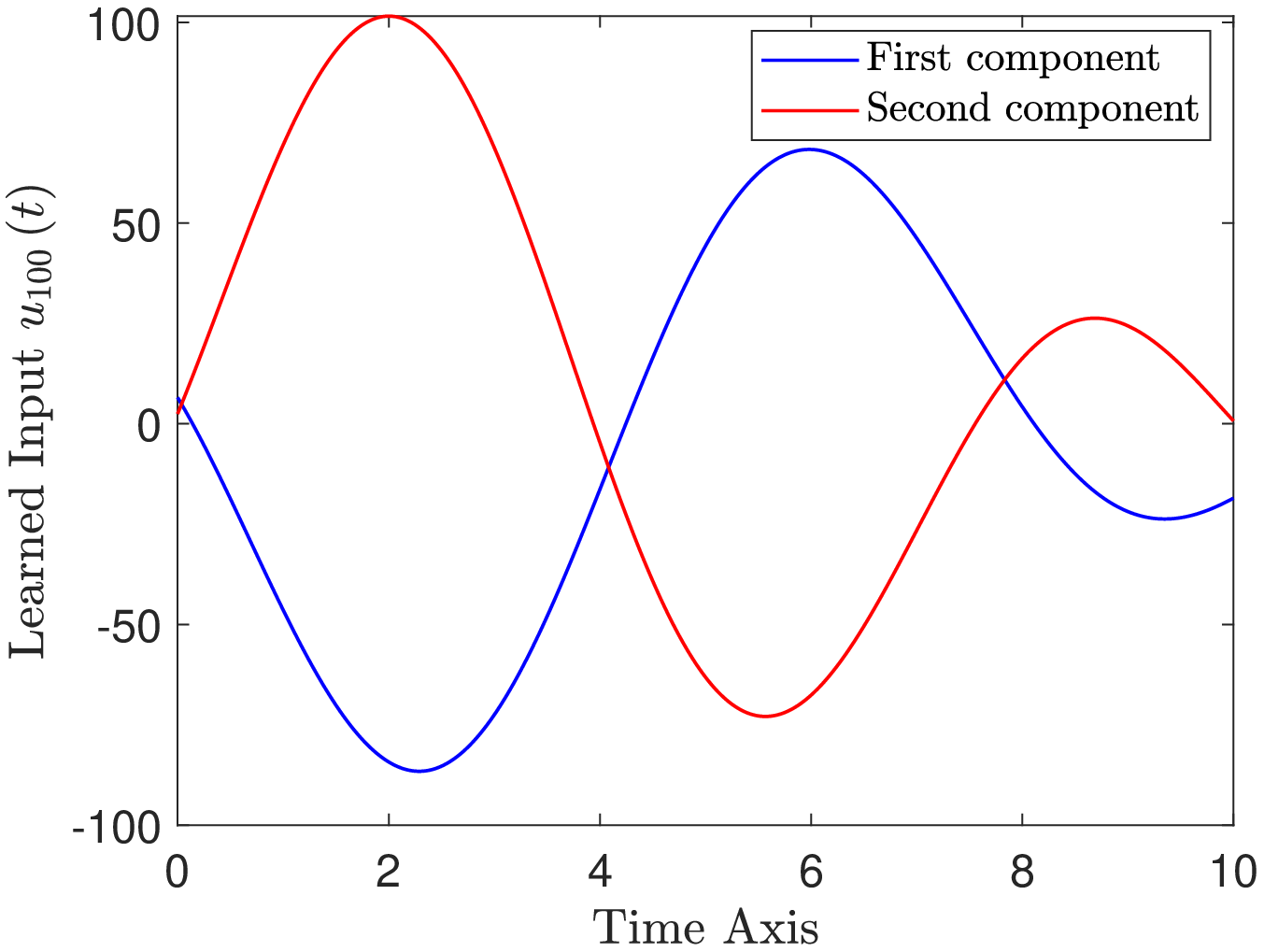}~\includegraphics[width=0.32\hsize]{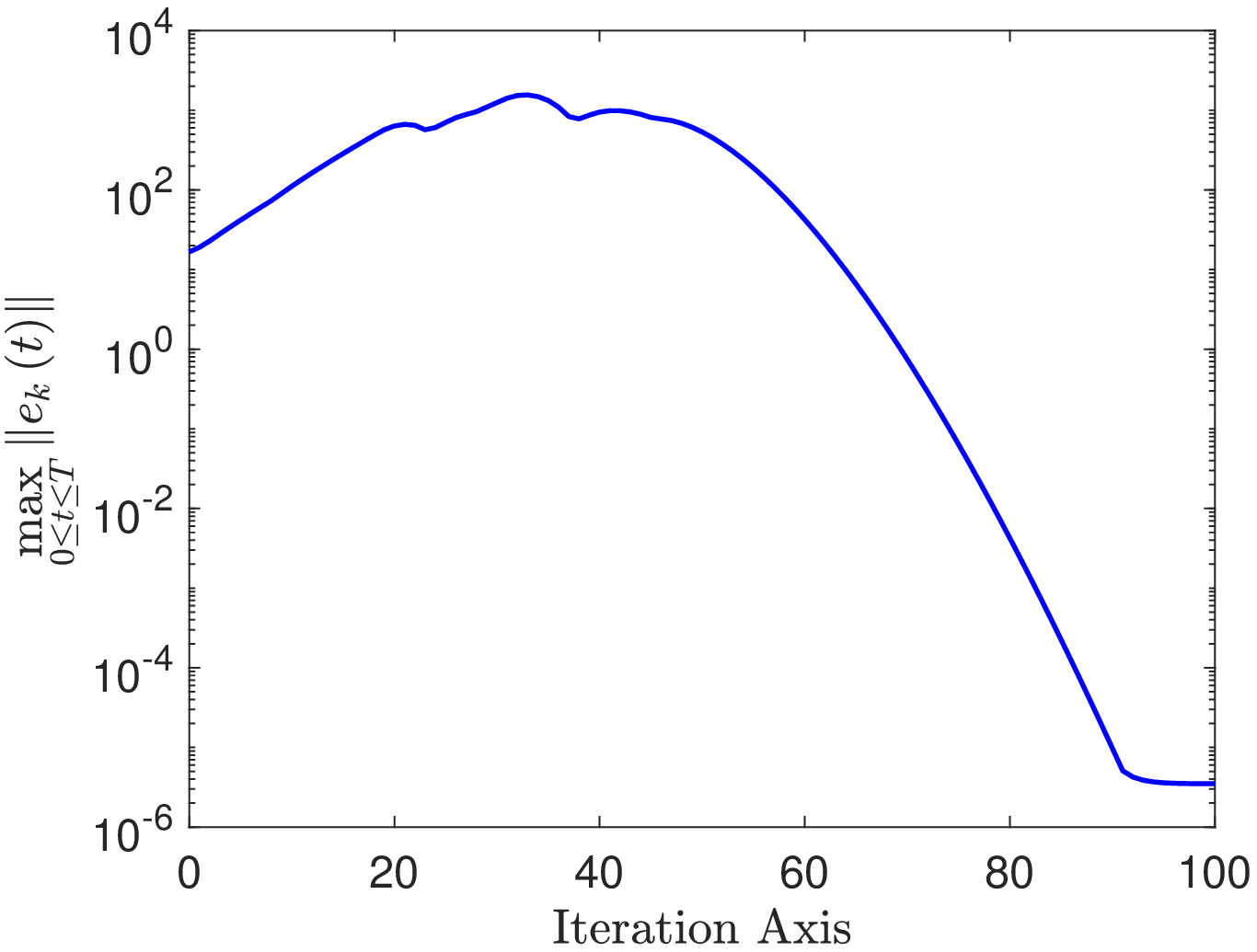}\\
\includegraphics[width=0.32\hsize]{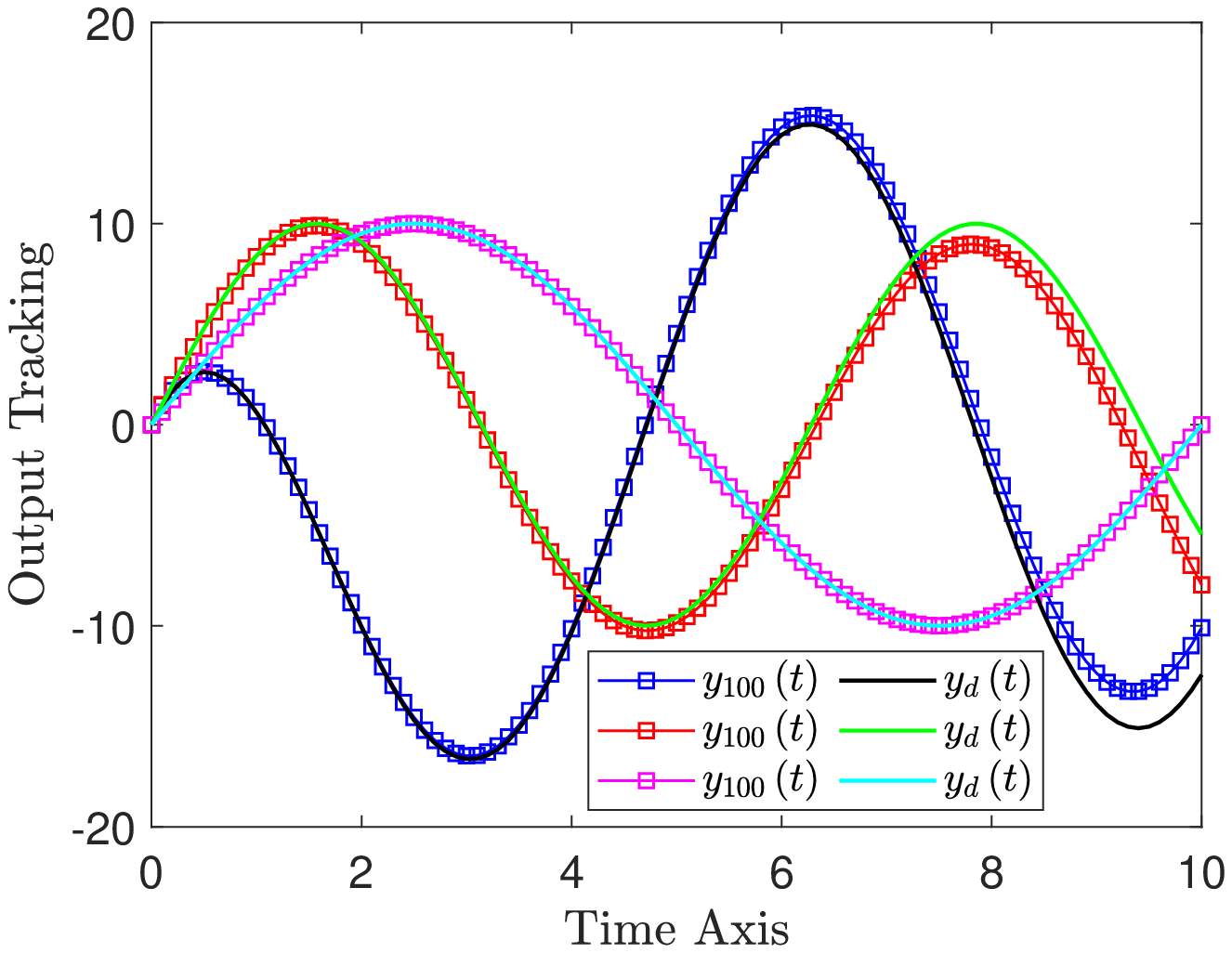}~\includegraphics[width=0.32\hsize]{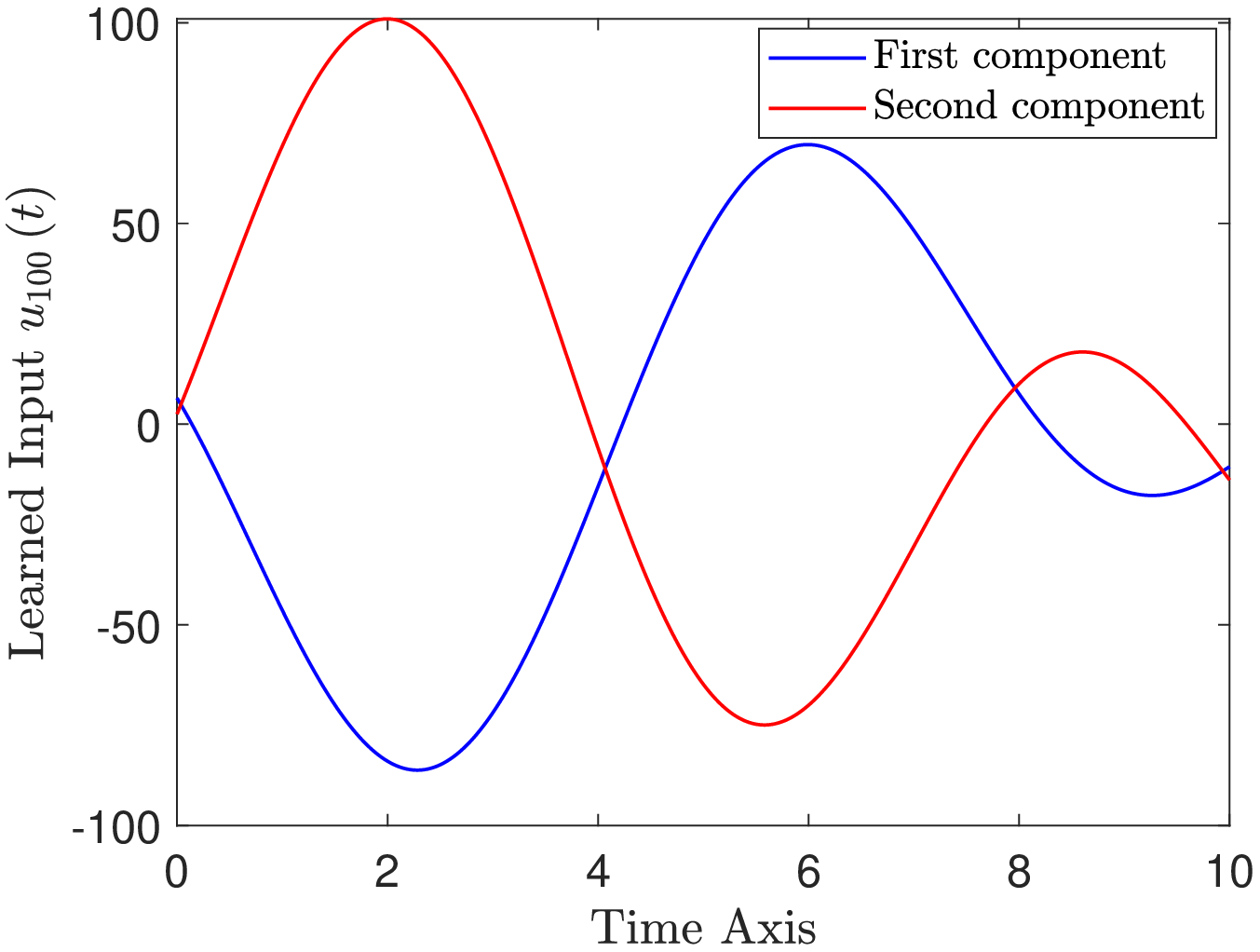}~\includegraphics[width=0.32\hsize]{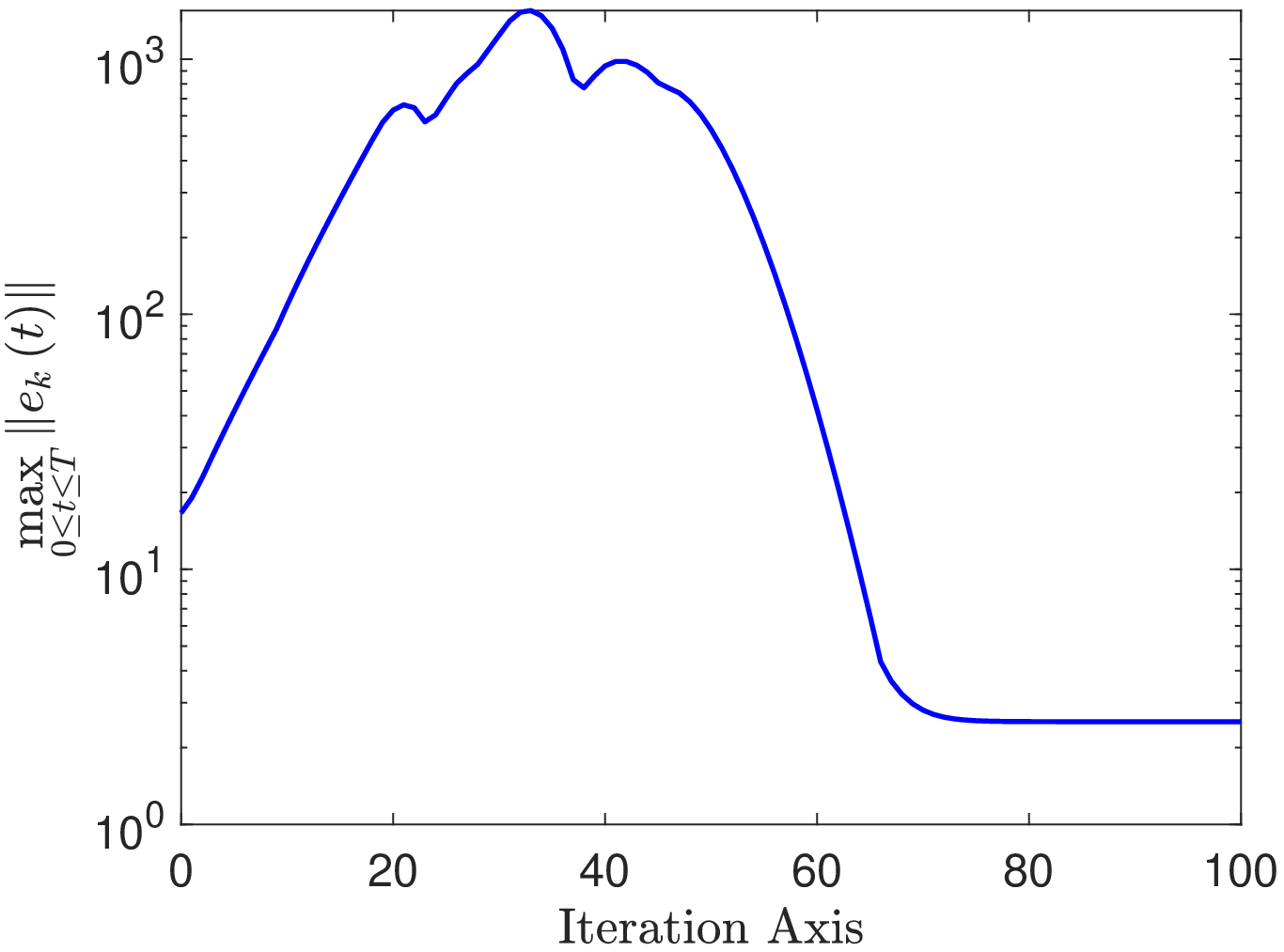}
\caption{(Example 1). Tracking performances of ILC for $q\geq p$. Upper: Case a). Middle: Case b). Lower: Case c).}
\label{fig01}
\end{figure*}

{\it Example 1:} Consider the system (\ref{b2}), where
\[
\aligned
G_1(s)
&=\begin{bmatrix}
\Ds \frac{12s-17}{10(s^2 + 3s + 2)} & \Ds \frac{12s-17}{10(s^2 + 3s + 2)} \\
\Ds \frac{111s + 82}{100(s^2 + 3s + 2)} & \Ds \frac{111s + 82}{100(s^2 + 3s + 2)} \\
\Ds \frac{25s^2-133s-216}{50(s^3 + 6s^2 + 11s + 6)} & \Ds \frac{61s^2 -25s - 144}{50(s^3 + 6s^2 + 11s + 6)}
\end{bmatrix}\\
G_2(s)
&=\begin{bmatrix}
\Ds \frac{1}{s+1} & \Ds \frac{1}{s+1} & \Ds \frac{s - 39}{10(s^2 + 3s + 2)} \\
\Ds \frac{1}{10(s+1)} & \Ds \frac{1}{10(s+1)} & \Ds \frac{5s + 3}{5(s^2 + 3s + 2)} \\
\Ds \frac{1}{5(s+1)} & \Ds \Ds \frac{5s+7}{5(s^2+4s+3)} & \Ds \frac{s^2 -20s - 29}{5(s^3 + 6s^2 + 11s + 6)}
\end{bmatrix}\\
D(s)&=0.
\endaligned
\]

\noindent We clearly have $q=3$ and $p=2$, and can verify the conditions C1)--C3). In addition, we consider three different cases for the specified output trajectory and the initial input as follows:
\[\aligned
\hbox{a)}~~y_d(t)
&=\begin{bmatrix}
\Ds\frac{57420}{3809}\left[\cos(t)-\exp\left(-\frac{82}{111}t\right)\right]-\frac{1240}{3809}\sin(t) \\
10\sin(t)\\
10\sin(5\pi/t)
\end{bmatrix}\\
u_0(t)&=0\\
\hbox{b)}~~y_d(t)
&=\begin{bmatrix}
\Ds\frac{57420}{3809}\left[\cos(t)-\exp\left(-\frac{82}{111}t\right)\right]-\frac{1240}{3809}\sin(t) \\
10\sin(t)\\
10\sin(5\pi/t)
\end{bmatrix}\\
u_0(t)
&=\begin{bmatrix}10 &-10\end{bmatrix}^{\tp}\\
\hbox{c)}~~y_d(t)
&=\begin{bmatrix}
\Ds\frac{57420}{3809}\left[\cos(t)-\exp\left(-\frac{82}{111}t\right)\right]-\frac{2}{5}\sin(t) \\
10\sin(t)\\
10\sin(5\pi/t)
\end{bmatrix}\\
u_0(t)&=0
\endaligned
\]

\noindent where, for all cases, we have $y_{d}(t)\in C^{1}_{3}[0,T]$, $u_{0}(t)\in C_{2}[0,T]$, and the initial condition (\ref{b7}). Because of $q>p$, we know from Theorem \ref{thm01} that $y_{d}(t)$ is trackable in ILC for the cases a) and b) since $Y_{d}(s)=\mathcal{L}\left[y_{d}(t)\right]$ satisfies the algebraic equation (\ref{b8}) for both cases, but it is not for the case c). To carry out simulations with the ILC updating law (\ref{a12}), we choose $\Gamma(s)$ as
\[
\Gamma(s)
=s\begin{bmatrix}
0.6849  &  0.6335  &  -1.25\\
-0.2807  &  -0.2596  &  1.25\\
\end{bmatrix}
\]

\noindent with which $\Gamma(s)G_{1}(s)$ is proper and satisfies (\ref{b13}).

Let $T=10$, and we plot the simulation results for the Cases a)--c) in Fig. \ref{fig01}. It is obvious from this figure that for the Cases a) and b), the tracking errors decrease to zero with increasing iterations, where the outputs learned after $100$ iterations track the specified trajectory perfectly. With the comparison between the learned input trajectories for the Cases a) and b) in Fig. \ref{fig01}, they are the same even though we adopt different initial inputs for both cases. This is consistent with the uniqueness result of Theorems \ref{thm01} and \ref{thm03} for the input that can generate the trackable trajectory in ILC. By contrast to the Cases a) and b), the Case c) considers a specified trajectory that is not trackable in ILC. Correspondingly, as depicted in Fig. \ref{fig01}, the tracking error does not decrease to zero with the increasing of iterations although the input still converges in the Case c), where, in particular, the output learned after $100$ iterations can no longer perfectly track the specified trajectory. By these observations, we demonstrate the trackability criterion of Theorem \ref{thm01} for the case $q\geq p$ and the relevant trackability-based ILC tracking result of Theorem \ref{thm03}, together with revealing the relation between the trackability of the specified output trajectory and the accomplishment of the perfect output tracking task in ILC.

\begin{figure*}[!t]
\centering
\includegraphics[width=0.32\hsize]{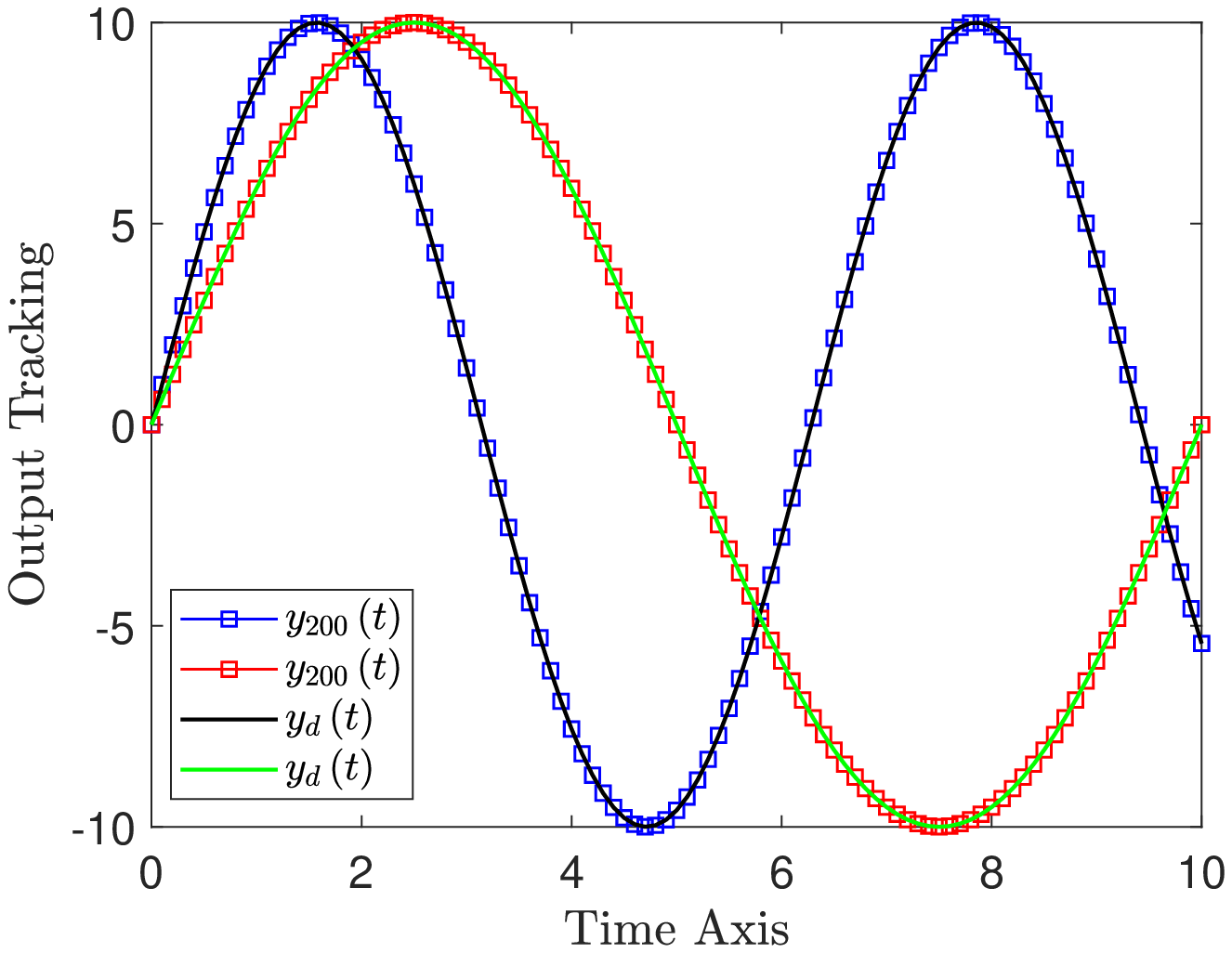}~\includegraphics[width=0.32\hsize]{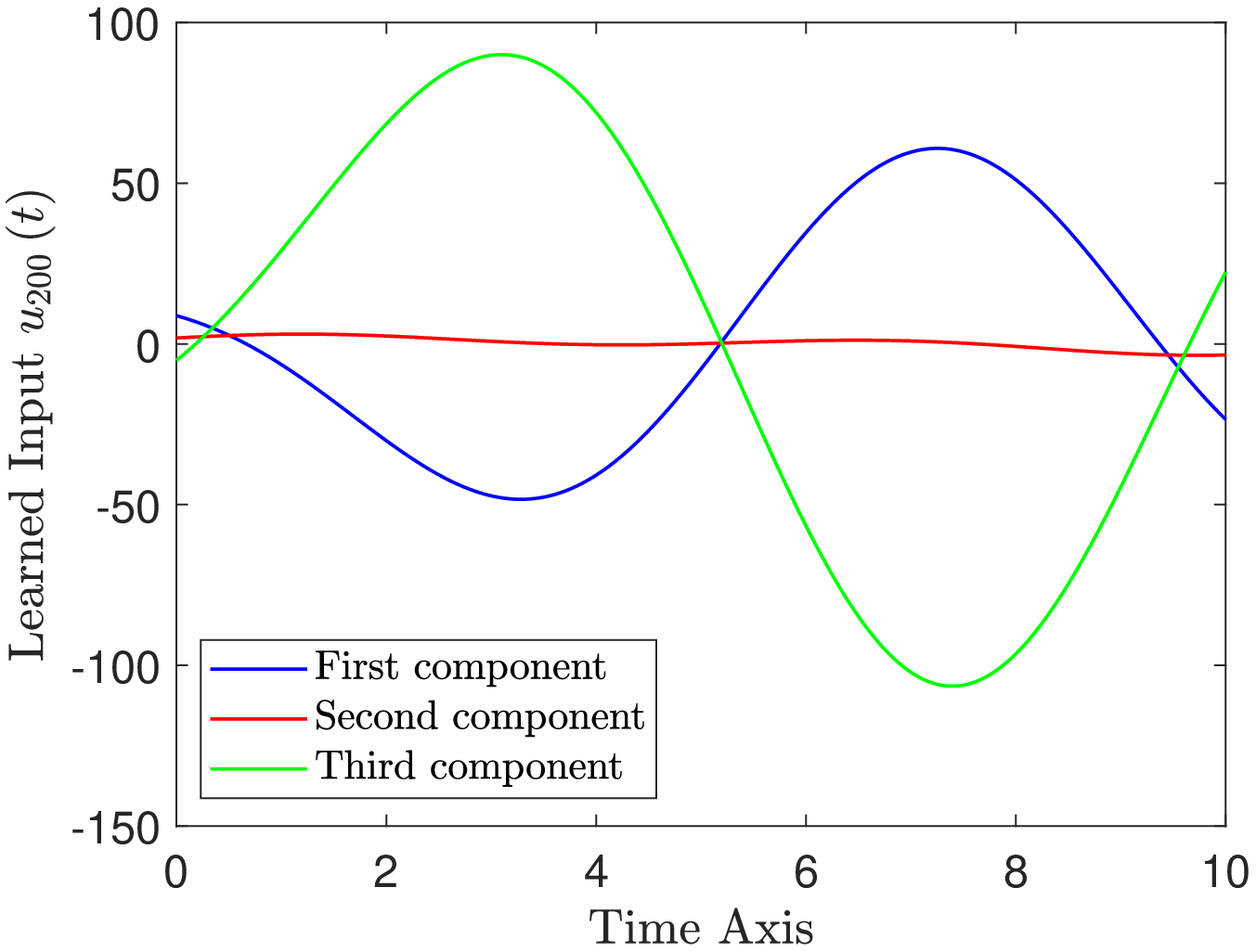}~\includegraphics[width=0.32\hsize]{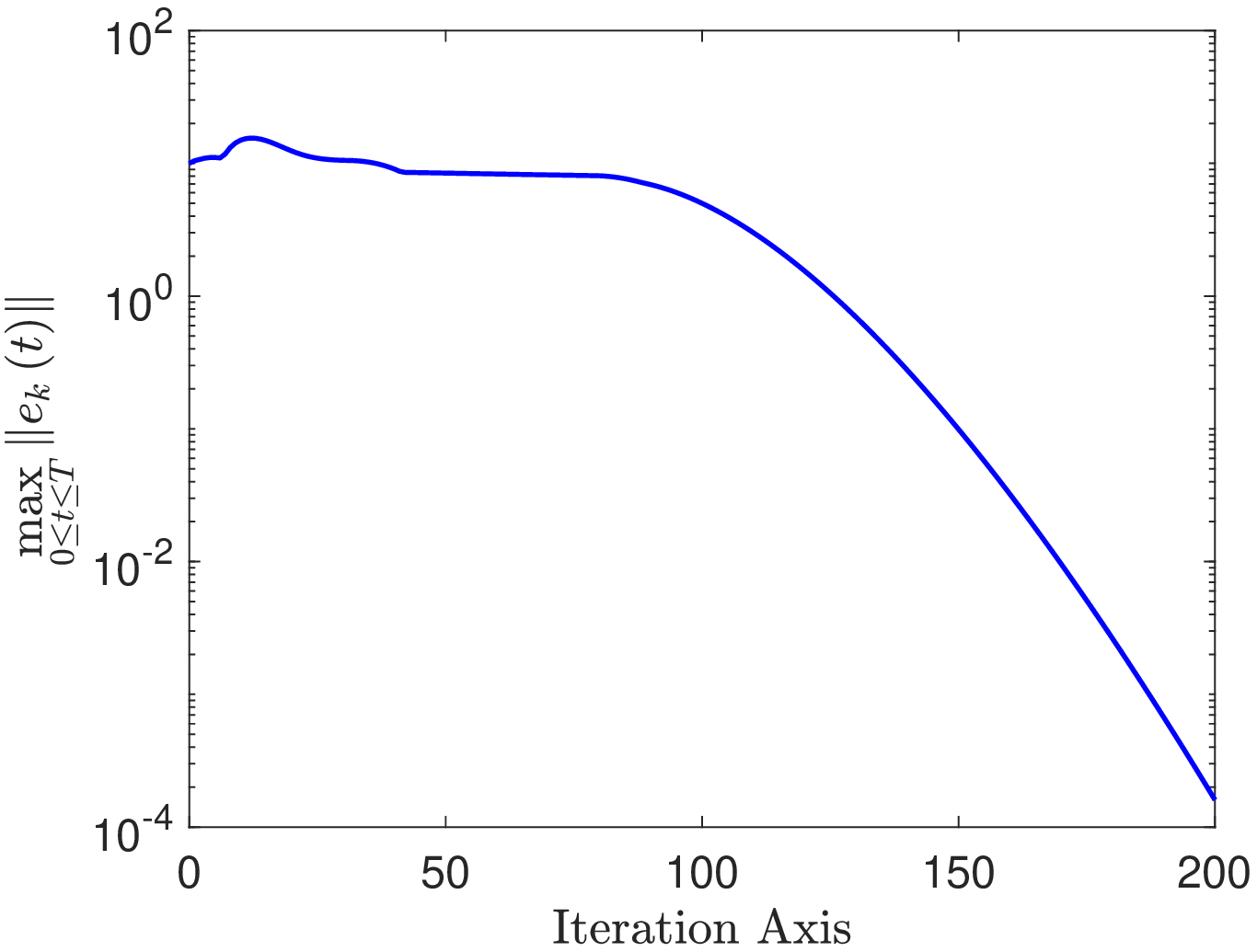}\\
\includegraphics[width=0.32\hsize]{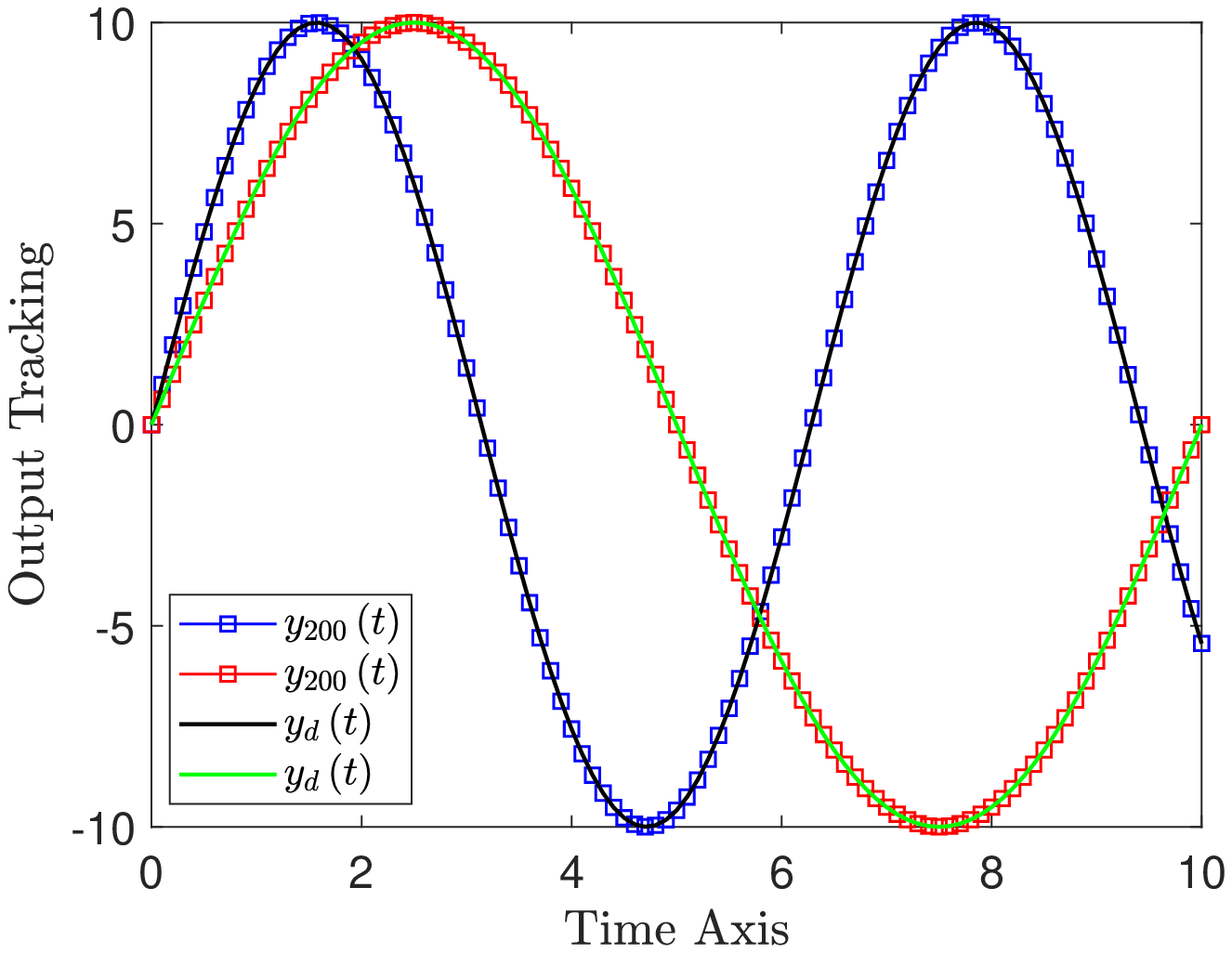}~\includegraphics[width=0.32\hsize]{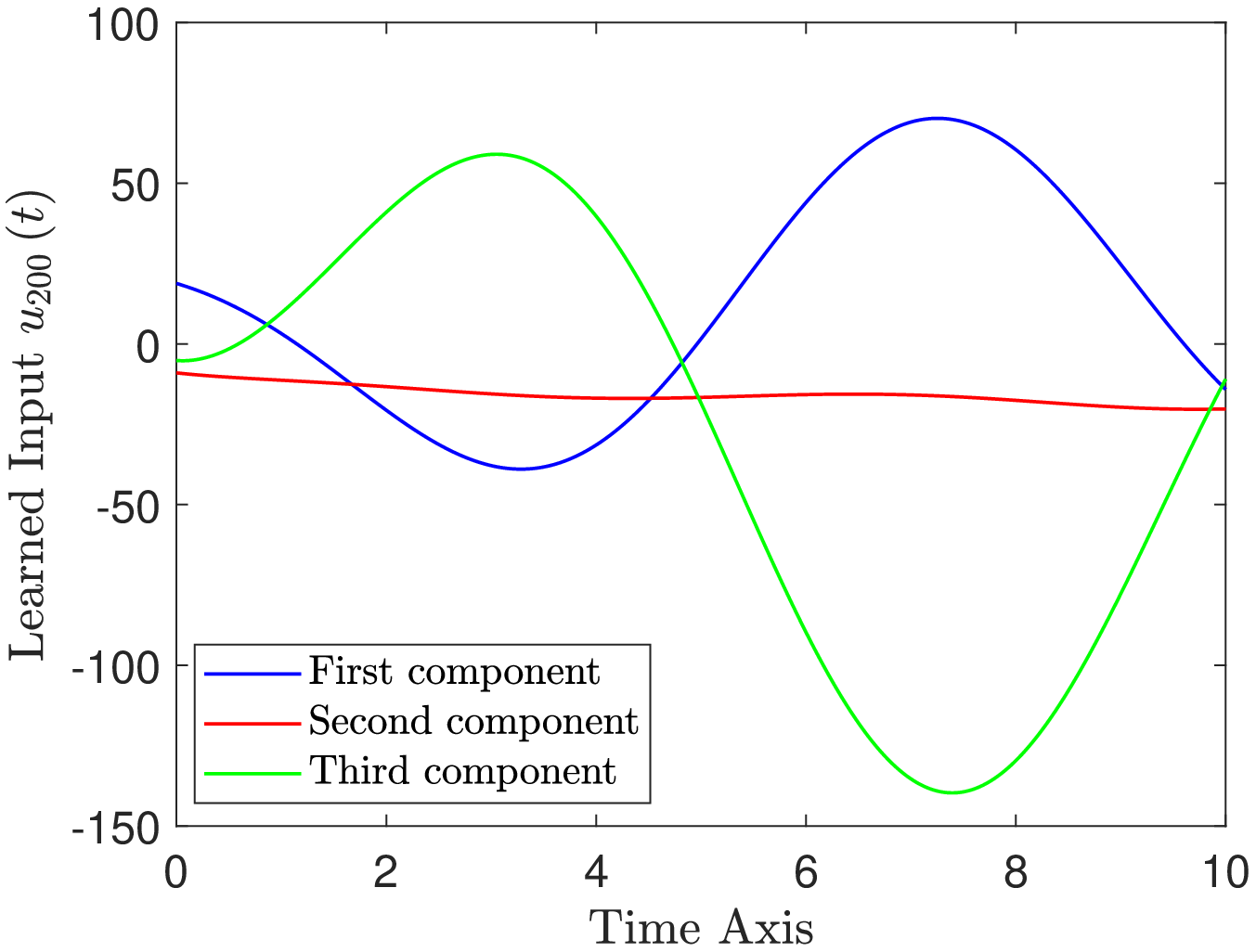}~\includegraphics[width=0.32\hsize]{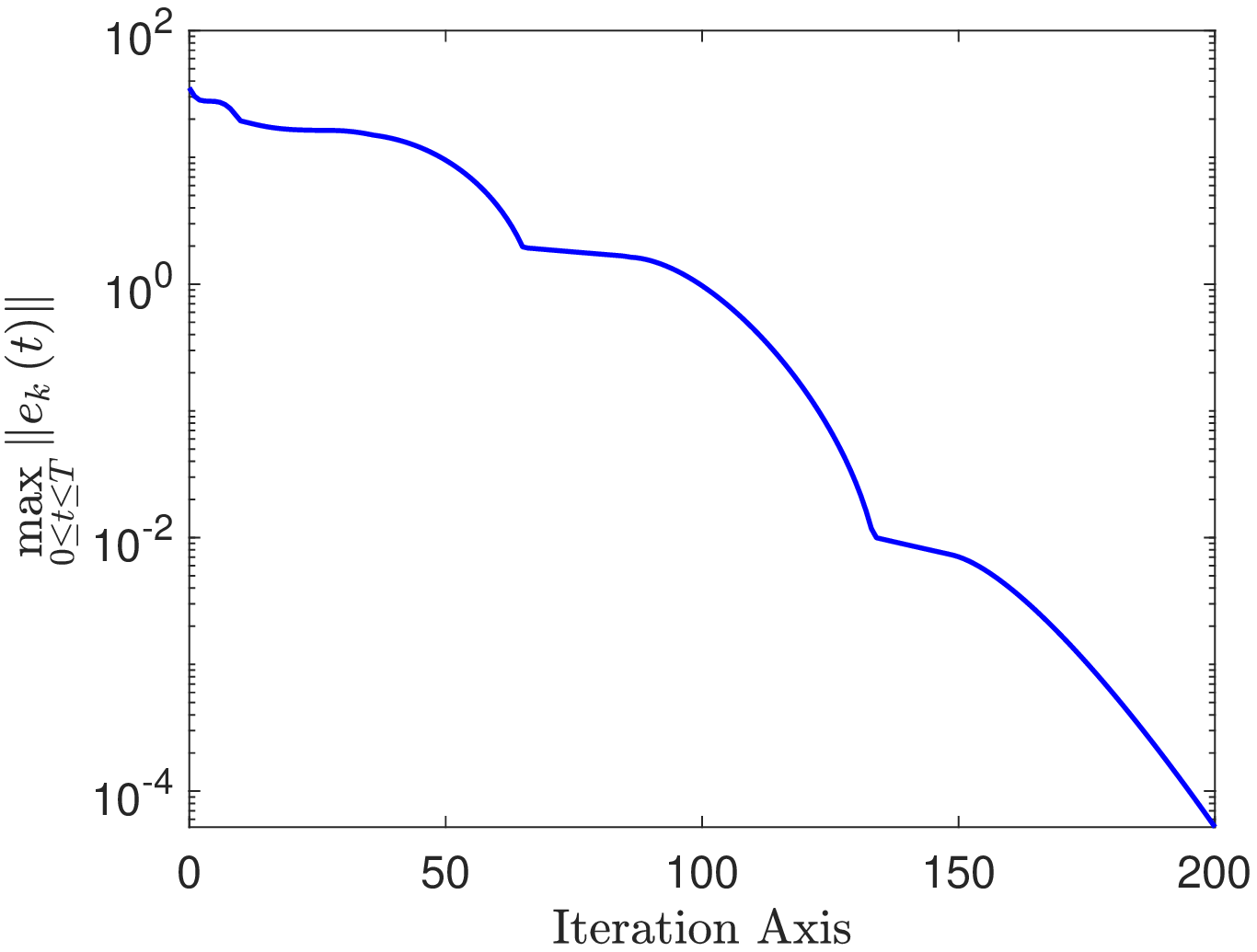}
\caption{(Example 2). Tracking performances of ILC for $q\leq p$. Upper: Case d). Lower: Case e).}
\label{fig02}
\end{figure*}

{\it Example 2:} Let $q=2$ and $p=3$, and then we consider the system (\ref{b2}) with
\[
\aligned
G_1(s)
&=\begin{bmatrix}
\Ds\frac{24s^2 + 155s + 195}{20(s^3 + 6s^2 + 11s + 6)}&\Ds\frac{120s^2 + 361s + 273}{100(s^3 + 6s^2 + 11s + 6)}\\
\Ds\frac{111s^2 + 523s - 348}{100(s^3 + 6s^2 + 11s + 6)}&\Ds\frac{111s^2 - 53s - 240}{100(s^3 + 6s^2 + 11s + 6)}\\
\Ds\frac{5s^2 + 42s + 45}{10(s^3 + 6s^2 + 11s + 6)}& \Ds\frac{61s^2 + 156s + 99}{50(s^3 + 6s^2 + 11s + 6)}
\end{bmatrix}^{\tp}\\
G_2(s)
&=\begin{bmatrix}
\Ds\frac{1}{s + 1}
&\Ds\frac{1}{10(s + 1)}\\
\Ds \frac{s+21}{10(s^2+4s+3)}
&\Ds \frac{5s+6}{5(s^2+4s+3)} \\
\Ds\frac{2s^2 + 9s - 9}{2(s^3 + 6s^2 + 11s + 6)}
&\Ds \frac{10s^2 - 9s - 27}{10(s^3 + 6s^2 + 11s + 6)}
\end{bmatrix}^{\tp}\\
D(s)&=0
\endaligned\]

\noindent for which the conditions C1)--C4) are satisfied. By taking $T=10$, we are interested in the following two cases of the specified output trajectory and the initial input:
\[\aligned
\hbox{d)}~~y_d(t)&=
\begin{bmatrix}
10\sin(t)\\
10\sin(\pi t/5)
\end{bmatrix},
\quad
u_{0}(t)=0\\ 
\hbox{e)}~~y_d(t)&=
\begin{bmatrix}
10\sin(t)\\
10\sin(\pi t/5)
\end{bmatrix},
\quad
u_{0}(t)=
\begin{bmatrix}10&-10&5\end{bmatrix}^{\tp}
\endaligned
\]

\noindent where $y_{d}(t)\in C^{1}_{2}[0,T]$ and $u_{0}(t)\in C_{3}[0,T]$. For both cases, we can also verify that the initial condition (\ref{b7}) holds, and therefore $y_{d}(t)$ is trackable in ILC by Theorem \ref{thm02}. To implement the ILC updating law (\ref{a12}), we select the gain matrix operator $\Gamma(s)$ as
\[
\Gamma(s)
=s\begin{bmatrix}
0.6 & 0.1 \\
0.1 & 0.1 \\
-0.5 & 0.4
\end{bmatrix}
\]

\noindent which makes $G_{1}(s)\Gamma(s)$ be proper such that (\ref{b20}) holds.

In Fig. \ref{fig02}, we depict the simulation results for both Cases d) and e), from which the zero convergence of the output tracking error along the iteration axis can be observed. In particular, for both cases, the outputs learned after $200$ iterations are capable of tracking the specified trajectory perfectly, despite which the input trajectories learned after $200$ iterations are different from each other since they correspond to different initial inputs. This validates not only the trackability-based ILC result of Theorem \ref{thm04}, but also the heavy dependence of the input learned with ILC on the initial input for the case $q\leq p$.

{\it Discussions:} By Examples 1 and 2, we illustrate the validity of our trackability-based ILC analysis for both under-actuated and over-actuated systems. It is clear that the trackability plays a crucial role in realizing the perfect ILC tracking task. Further, Figs. \ref{fig01} and \ref{fig02} demonstrate that it is feasible to employ a unified condition to implement the ILC convergence analysis from the perspectives of both input and output.

\section{Conclusions}\label{sec5}

In this paper, we have discussed the fundamental trackability problems for continuous-time ILC systems. We have explored the trackability criteria with the help of utilizing the frequency-domain algebraic equations to determine whether the specified output trajectory is trackable in ILC, despite under-actuated or over-actuated systems. In particular, we have investigated how to arrive at all inputs that can generate the trackable trajectory. The (uniform) convergence analysis has been implemented by newly developing an FCS-induced method of ILC. It has been disclosed that the perfect output tracking task of ILC is closely connected to the trackability of the specified output trajectory. Our proposed trackability-based ILC results have been verified through two simulation examples.


\appendix

\begin{IEEEproof}[Proof of Lemma \ref{lem08}]
By Definition \ref{defi1} and with \cite[Chapter 1, Theorem 5.3]{am:06}, we directly have the equivalence between 1) and 2). Next, we prove the equivalence between 1) and 3).

{\it Sufficiency:} Because $G_F(s)$ is proper, let us denote
\begin{equation}\label{a55}
D_{F}=\lim_{s\to\infty}G_{F}(s).
\end{equation}

\noindent Based on (\ref{a55}) and thanks to $\rho\left(\lim_{s\to\infty}G_{F}(s)\right)<1$, there exists some induced matrix norm \cite{hj:85} such that
\begin{equation}
\label{a56}
\|D_{F}\|\leq\rho_1
\end{equation}

\noindent where $0\leq\rho_1<1$. By (\ref{a55}), we write $G_F(s)$ in the form of
\begin{equation}\label{a54}
G_F(s)=\widehat{G}_F(s)+D_{F}
\end{equation}

\noindent and thus $\widehat{G}_F(s)\in\R\F^{n\times n}(s)$ is strictly proper such that $\Phi_{F}(t)\triangleq\mathcal{L}^{-1}\left[\widehat{G}_F(s)\right]\in\R^{n\times n}$ is smooth. Let $\beta_{F}\triangleq\max_{t\in[0,T]}\left\|\Phi_{F}(t)\right\|$, and it is obvious that $\beta_{F}$ is finite. By incorporating (\ref{a54}), we can get from $F_{k+2}(s)-F_{k+1}(s)=G_{F}(s)\left[F_{k+1}(s)-F_{k}(s)\right]$, $\forall k\in\mathbb{Z}_{+}$ that, for all $t\in[0,T]$ and for all $k\in\mathbb{Z}_{+}$,
\begin{equation}\label{a58}
\aligned
f_{k+2}(t)-f_{k+1}(t)&=\int_{0}^{t}\Phi_{F}(t-\tau)\left[f_{k+1}(\tau)-f_{k}(\tau)\right]d\tau\\
&~~~+D_{F}\left[f_{k+1}(t)-f_{k}(t)\right].
\endaligned
\end{equation}

\noindent Then by taking the norm on both sides of (\ref{a58}) and leveraging (\ref{a56}), we can arrive at
\begin{equation}
\label{a59}
\aligned
\left\|f_{k+2}(t)-f_{k+1}(t)\right\|
&\leq\int_{0}^{t}\left\|\Phi_{F}(t-\tau)\right\|\left\|f_{k+1}(\tau)-f_{k}(\tau)\right\|d\tau\\
&~~~+\left\|D_{F}\right\|\left\|f_{k+1}(t)-f_{k}(t)\right\|\\
&\leq\beta_{F}\int_{0}^{t}\left\|f_{k+1}(\tau)-f_{k}(\tau)\right\|d\tau\\
&~~~+\rho_{1}\left\|f_{k+1}(t)-f_{k}(t)\right\|
\endaligned
\end{equation}

\noindent for which we consider any $\lambda>0$ and can verify
\begin{equation}
\label{a61}
\aligned
\int_{0}^{t}\|f_{k+1}(\tau)-f_{k}(\tau)\|d\tau
&=\int_{0}^{t}e^{\lambda\tau}\left[e^{-\lambda \tau}\|f_{k+1}(\tau)-f_{k}(\tau)\|\right]d\tau\\
&\leq\|f_{k+1}(t)-f_{k}(t)\|_{\lambda}\int_{0}^{t}e^{\lambda \tau}d\tau\\
&=\frac{\Ds e^{\lambda t}-1}{\Ds\lambda}\|f_{k+1}(t)-f_{k}(t)\|_{\lambda}.
\endaligned
\end{equation}

\noindent To proceed with (\ref{a59}) and (\ref{a61}), we can further obtain
\begin{equation*}\label{a60}
\aligned
e^{-\lambda t}\|f_{k+2}(t)-f_{k+1}(t)\|
&\leq\beta_{F}e^{-\lambda t}\int_{0}^{t}\|f_{k+1}(\tau)-f_{k}(\tau)\|d\tau\\
&~~~+\rho_1e^{-\lambda t}\|f_{k+1}(t)-f_{k}(t)\|\\
&\leq\beta_{F}\frac{\Ds1-e^{-\lambda t}}{\Ds\lambda}\|f_{k+1}(t)-f_{k}(t)\|_{\lambda}\\
&~~~+\rho_1e^{-\lambda t}\|f_{k+1}(t)-f_{k}(t)\|\\
&\leq\left(\rho_{1}+\lambda^{-1}\beta_{F}\right)\|f_{k+1}(t)-f_{k}(t)\|_{\lambda}
\endaligned
\end{equation*}

\noindent which implies that, for all $k\in\mathbb{Z}_{+}$,
\begin{equation}\label{a63}
\|f_{k+2}(t)-f_{k+1}(t)\|_{\lambda}
\leq\left(\rho_{1}+\lambda^{-1}\beta_{F}\right)\|f_{k+1}(t)-f_{k}(t)\|_{\lambda}.
\end{equation}

\noindent Thanks to $\rho_{1}\in[0,1)$, we choose $\lambda>0$ such that $\rho_{1}+\lambda^{-1}\beta_{F}\leq\left(\rho_{1}+1\right)/2\triangleq\rho$. Clearly, $\rho\in[0,1)$ holds, and consequently, the use of (\ref{a63}) results in
\[
\|f_{k+2}(t)-f_{k+1}(t)\|_{\lambda}
\leq\rho\|f_{k+1}(t)-f_{k}(t)\|_{\lambda},\quad\forall k\in\mathbb{Z}_{+}
\]

\noindent by which we have
\begin{equation}\label{a64}
\|f_{k+1}(t)-f_{k}(t)\|_{\lambda}
\leq\rho^{k}\|f_{1}(t)-f_{0}(t)\|_{\lambda},\quad\forall k\in\mathbb{Z}_{+}.
\end{equation}

\noindent Then from (\ref{a64}), we know that for any $\epsilon>0$, there exists some integer $N(\epsilon)\geq\max\{0,\ln(\epsilon (1-\rho)/\|f_{1}(t)-f_{0}(t)\|_{\lambda})/\ln(\rho)\}$ such that 
\[
\aligned
\|f_{i}(t)-f_{j}(t)\|_{\lambda}&\leq \sum_{k=j}^{i-1} \|f_{k+1}(t)-f_{k}(t)\|_{\lambda}\\
&\leq \sum_{k=j}^{i-1} \rho^k \|f_{1}(t)-f_{0}(t)\|_{\lambda}\\
&\leq \sum_{k=N(\epsilon)}^{\infty} \rho^k \|f_{1}(t)-f_{0}(t)\|_{\lambda}\\
&\leq \frac{\rho^{N(\epsilon)}}{1-\rho}\|f_{1}(t)-f_{0}(t)\|_{\lambda}\\
&\leq \epsilon,\quad\forall i\geq j\geq N(\epsilon).
\endaligned
\]

\noindent Similarly, we can also get $\|f_{i}(t)-f_{j}(t)\|_{\lambda}\leq\epsilon$, $\forall j\geq i\geq N(\epsilon)$. Then from Definition \ref{defi1}, it follows that the functional sequence $\{f_k(t): k \in \Z_+\}$ is an FCS.

{\it Necessity:} If $\{f_k(t): k \in \Z_+\}$ is an FCS, then with Definition \ref{defi1}, $\lim_{k\to\infty}\Delta f_{k}(t)=0$, $\forall t\in[0,T]$ holds. For any given $t\in[0,T]$, this is actually an asymptotic stability result of the system (\ref{a58}) along the iteration axis since (\ref{a58}) essentially denotes a discrete linear system given by
\begin{equation*}\label{}
\Delta f_{k+1}(t)
=D_{F}\Delta f_{k}(t)+\int_{0}^{t}\Phi_{F}(t-\tau)\Delta f_{k}(\tau)d\tau,\quad\forall k\in\mathbb{Z}_{+}.
\end{equation*}

\noindent It basically requires $\rho\left(D_{F}\right)<1$, that is, $\rho\left(\lim_{s\to\infty}G_{F}(s)\right)<1$. For the necessity of Lemma \ref{lem08}, the readers can also be referred to that of \cite[Lemma 1]{lch:05} because (\ref{a58}) can be described in the form of the 2-D linear continuous-discrete system \cite[(13)]{lch:05}.

Furthermore, if $f_{k}(t)\in C_{n}[0,T]$, $\forall k\in\mathbb{Z}_{+}$, then we can easily conclude from the completeness of the space $C_{n}[0,T]$ that there exists some function $f_{\infty}(t)\in C_{n}[0,T]$ such that $\lim_{k \to \infty} \|f_k(t)-f_{\infty}(t)\|_{\lambda}=0$, i.e., $\lim_{k \to \infty}f_k(t)=f_{\infty}(t)\in C_{n}[0,T]$.
\end{IEEEproof}

\begin{IEEEproof}[Proof of Lemma \ref{lem07}]
Let $\overline{\Phi}(t)\triangleq\mathcal{L}^{-1}\left[\overline{G}(s)\right]$. Since $\overline{G}(s)$ is proper, $\overline{\Phi}(t)$ satisfies $\overline{\Phi}(t)=\overline{\Phi}_{sp}(t)+D_{\overline{G}}\delta(t)$, where $\overline{\Phi}_{sp}(t)$ is smooth and $D_{\overline{G}}\in\mathbb{R}^{m\times n}$ is such that $\lim_{s\to\infty}\overline{G}(s)=D_{\overline{G}}$. By noticing $\overline{f}(t)=\mathcal{L}^{-1}\left[\overline{G}(s)F(s)\right]$, we can validate
\[
\overline{f}(t)=\int_{0}^{t}\overline{\Phi}_{sp}(t-\tau)f(\tau)d\tau+D_{\overline{G}}f(t), \quad t\in[0,T]
\]

\noindent from which $\overline{f}(t)\in C_{m}[0,T]$ is immediate due to $f(t)\in C_{n}[0,T]$. Moreover, if $\overline{G}(s)$ is strictly proper, then $\lim_{s\to\infty}\overline{G}(s)=0$ and, consequently, $\overline{\Phi}(t)=\mathcal{L}^{-1}\left[\overline{G}(s)\right]$ is smooth, which guarantees $\mathcal{L}^{-1}\left[\overline{G}(s)v\right]=\overline{\Phi}(t)v\in C_{m}[0,T]$ for any $v\in\mathbb{R}^{n}$.
\end{IEEEproof}

\bibliographystyle{ieeetr}

\end{document}